# Solar System Dynamics and Multiyear Droughts of the Western USA

James H. Shirley

*TORQUEFX*

Simi Valley, California, USA

jrocksci@att.net

orcid.org/000-001-6496-8975

3 December 2021

Key Points:

- We calculate the time history of orbit-spin coupling torques on the Earth system for the interval from 1860-2040
- A one-to-one relationship links extended periods of reduced torque amplitude with multiyear droughts recorded in the Western USA since 1860
- A future multiyear episode of widespread drought in the Western USA is forecasted to begin in 2028 ± 4 yr (2 σ)

**Abstract:**

      The recent addition of orbit-spin coupling torques to atmospheric global circulation models has enabled successful years-in-advance forecasts of global and regional-scale dust storms on Mars. Here we explore the applicability of the orbit-spin coupling mechanism for understanding and forecasting anomalous weather and climate events on Earth. We calculate the time history of orbit-spin coupling torques on the Earth system for the interval from 1860-2040. The torque exhibits substantial variability on decadal to bidecadal timescales. Deep minima recur at intervals from 15-26 years; eight such episodes are documented within the study period prior to 2020. Each of the identified torque minima corresponds in time to an episode of widespread drought in the Western USA extending over several years. The multiyear droughts of the 1930's, the 1950's, the mid-1970's, the early 1990's, and of 2011-2015 were each coincident in time with orbit-spin coupling torque minima. The upcoming torque minimum of 2030 is the deepest such minimum of the 180-yr study interval. A multiyear episode of widespread drought in the Western USA is likely to be underway by 2028 ± 4 years (2 σ). The potential benefits to societies of improved drought predictions justify an immediate high-priority effort to include forcing by orbit-spin coupling within state-of-the-art Earth system GCMs. Future targeted numerical modeling investigations are likely to yield forecasts with considerably lower uncertainties and with much improved temporal resolution in comparison to that obtained here.

# 1. Introduction

Multiyear episodes of widespread drought in the Western USA (Cook et al. 1997, 2004, 2007; Herweijer et al., 2006; Parsons & Coats, 2019) exhibit statistically significant recurrence tendencies on bidecadal timescales (Mitchell et al., 1979; Currie, 1981, 1984a, 1984b; Bell, 1981a, 1981b; Stockton et al., 1983; Cook et al., 1997). Two independent signatures are frequently found; one of these has been characterized as a "22-yr rhythm" of drought. This cycle has been tentatively linked with the phasing of the Hale (double) sunspot cycle (Mitchell et al., 1979; Cook et al., 1997). The second signature corresponds to the 18.6-yr period of the retrograde revolution of the nodes of the Moon's orbit on the Earth's equatorial plane (Currie, 1981, 1984a, 1984b; Bell, 1981a, 1981b). Cook et al. (1997) employ multiple techniques to show that the two cycles operate independently, with additive effects. These authors conclude that "the statistical evidence for these linkages appears to be strong enough to justify the continued search for a physical model." Additional motivation for physical model development may be found in the large catalog of economic, social, and environmental costs of past multiyear drought episodes of the Western USA. Crop losses from a single such event may easily run into the billions or tens of billions of US dollars (Riebesame et al., 1991; Willhite, 2000; Ross & Lott, 2003; Basara et al., 2013). Drought triggered water shortages impact ecological, agricultural, and urban elements of societies. Soil moisture changes and aridity associated with prolonged droughts have significant impacts on natural ecosystems (Lotsch et al. 2005; He et al., 2014; Crausbay et al., 2017); measurable impacts on human health are likewise reported (Riebesame et al., 1991; Basara et al., 2013).

Physical explanations put forward to account for the 18.6-yr and 22-yr signals in prolonged droughts and in other atmospheric and oceanic indices are generally faulted on grounds of *quantitative insufficiency*. Solar irradiance variations over the 22-yr Hale activity cycle are much too small to effect significant changes in the large-scale circulation of the atmosphere, or in sea surface temperatures, which have long been considered likely to play a role in Western USA drought occurrence (Namias, 1978; Trenberth et al., 1988; Cook et al., 2007; Sung et al., 2014; Hartmann, 2015). Likewise, while a small (theoretically ≤ 15 mm) ocean tide is associated with the 18.6-yr revolution of the Moon's nodes (Ray, 2007; Cherniawsky et al., 2010), the associated tidal accelerations are demonstrably too small to produce effects within the atmosphere and oceans sufficient to explain the bidecadal signatures observed.

In this paper we consider for the first time the possible role of orbit-spin coupling (Shirley, 2017a; Mischna & Shirley, 2017; Shirley, Newman et al., 2019; Newman et al., 2019; Shirley, McKim et al. 2020) as a candidate mechanism for exciting Earth system variability on bidecadal timescales. We focus on the occurrence of widespread multiyear drought episodes in the Western USA, but many other terrestrial indices and observations are mentioned in passing. In Section 2 we provide an overview of the orbit-spin coupling hypothesis, along with a scale analysis demonstrating quantitative sufficiency. We calculate the orbit-spin coupling torques at intervals of 2 hours for the period from 1860-2040. We describe the variability of the torque on semidiurnal to annual timescales in Section 3. Variability on decadal to centennial timescales is described in Section 4. We identify separate solar system dynamical candidate sources for the 18.6-yr and ~22-yr components of variability.

Section 5 compares the historic record of multiyear Western USA drought episodes with calculated torque magnitudes since 1860. Deep minima of the torque occurred in 1873, 1891, 1910, 1929, 1951, 1975, 1990, and in 2013. A one-to-one correspondence is found linking each of these minima with an accompanying multiyear episode of widespread drought in the Western USA. The droughts comprising this pattern occurred in 1870-1877, 1890-1896, 1910-1912, 1929-1940, 1946-1956, 1975-1977, 1987-1992, and 2011-2015. We introduce a physical hypothesis for multiyear drought occurrence that may be unambiguously validated or falsified through numerical modeling.

Pertinent coupled ocean-atmosphere effects are addressed in Section 6, focusing principally on effects associated with the Pacific Decadal Oscillation (PDO) and its links to shorter period oscillations including El Nino and the Southern Oscillation (ENSO). Many published studies link western USA drought occurrence with the variability of sea surface temperatures (cf. Cook et al., 2007). We review this topic in some detail, as no successful physical model for drought occurrence may ignore the coupling of the ocean and atmosphere.

We assess the utility of our results for forecasting the near-term course of future multiyear Western USA drought occurrence in Section 7. Bearing in mind the limitations of our approach, we are nonetheless by statistical methods able to forecast the onset of a future multi-year episode of widespread drought in the Western USA in 2028 ± 4 yr.

In Section 8 we briefly discuss some wider implications of the new results presented in Sections 3, 4, and 5. Implications of our results with respect to the issue of greenhouse gas forcing of climate change are addressed in this Section. Section 8 also includes recommendations for future work. Results are summarized, and conclusions detailed, in Section 9.

**2. Physics of Atmospheric Motions and Orbit-Spin Coupling**

2.1. An overview

Prior investigations and publications describing the orbit-spin coupling hypothesis and detailing its effects have focused exclusively on the Mars atmosphere; atmospheric scientists working on terrestrial problems are thus unlikely to be familiar with the key concepts or with the extensive testing that has already been performed. Here we briefly introduce the physical mechanism and its predictions, describe the hypothesis testing that has been performed to date, and review the agreement with Martian atmospheric observations.

The orbit-spin coupling hypothesis posits and quantifies an exchange of angular momentum between the reservoirs of the planetary orbital motion and the rotational motion. The transfer is accomplished by a reversing torque whose axis lies within the equatorial plane of the planet (Section 2.2). The atmosphere participates in the exchange of momentum between the orbital and rotational reservoirs. Particles and volumes (or parcels) of the affected atmosphere gain and lose momentum (and velocity) in this process. In consequence, the large-scale circulation of the atmosphere is intermittently spun up, and subsequently spun down, over cycle times that are largely unrelated to (and incommensurate with) the known seasonal cycles driven by solar input to the atmospheric system. Mars general circulation model (MGCM) experiments (Mischna & Shirley, 2017; Newman et al., 2019) obtained occasional changes of up to ~20% in global mean surface wind speeds as a consequence of the coupling.

Spectacular global dust storms on Mars occur intermittently, sometimes after gaps of multiple Mars years. They are in that way similar to multiyear episodes of widespread drought on Earth. Occurrence times of future Martian global dust storms have long been considered to be unpredictable. However, the Martian global dust storm of 2018, which occurred five Mars years after the previous such storm, was successfully forecasted several years in advance, on the basis of solar system dynamical considerations (Shirley, 2015), via theoretical and statistical considerations (Shirley & Mischna, 2017), and on the basis of numerical global circulation modeling with orbit-spin coupling (Mischna & Shirley, 2017).

Outcomes of century-long MGCM simulations have been compared with the available historic record of Martian global dust storm occurrence. In the initial study (Mischna & Shirley, 2017), the forced version of the MGCM produced atmospheric wind systems favorable for Martian global dust storm occurrence *in the years and seasons in which such storms actually occurred*. In a subsequent, more elaborate study (Newman et al., 2019), modeling outcomes were found to reproduce the historic record of years with (and without) global storms with a success rate >78% (Shirley, Newman, et al., 2019). No data assimilation, or tuning of initial conditions, was required in either study. Large ensembles of model runs were likewise not required. Both of the cited numerical modeling investigations assert rigorous proof of concept for the physical hypothesis.

Time-varying meridional accelerations (directed alternately northward and southward) are in large part responsible for the resulting cascading changes in the circulation of the affected atmosphere (Shirley, 2017a, Section 4). The theoretically predicted intermittent intensification of meridional overturning circulations was initially confirmed via comparisons of forced- and control-model runs in century-long MGCM simulations (Mischna & Shirley, 2017). An anomalous powering-up of the Martian meridional overturning circulation, as simulated, was later observed directly, by orbiting spacecraft, during the earliest stages of the Martian global dust storm of 2018 (Shirley, Kleinböhl, et al., 2019).

A subsequent investigation (Shirley, McKim, et al., 2020) reveals that Martian global dust storms initiate 1) near times when the orbit-spin coupling torques are peaking, and 2) near times when the torques are changing most rapidly. Sub-seasonal timescale forecasts for future episodes of large-scale atmospheric instability on Mars are now available for the interval extending from 2020-2030 (Shirley, McKim, et al., 2020). A discussion of the correspondence between the forecasted torque episodes and large Martian regional dust storms occurring during the initial period of the above forecast interval may be found in Shirley, Kleinböhl, et al. (2020).

The atmospheres and atmospheric circulations of the Earth and Mars are remarkably similar in some ways, while they are considerably different in others. We list and discuss key similarities and differences in Appendix 1.

2.2. Governing equations

Equations for the conservation of energy, the conservation of mass, and the conservation of momentum comprise the fundamental *governing equations* describing terrestrial planet atmospheres in motion. We will be concerned with the momentum equation, which may be written as follows (Holton, 1992):

$$\frac{DU}{Dt} = -2\,\omega_\alpha \times U - \frac{1}{\rho} \nabla p + g + F_r \tag{1}$$

The forces represented on the right side are the Coriolis force, the pressure gradient force, effective gravity (including the centrifugal force), and the friction force. (Throughout this paper, the vector angular velocity of the planetary rotation, as required for the Coriolis force, will be designated by $\omega_\alpha$, after (Shirley, 2017a), rather than by $\Omega$, as in Holton (1992).

Global circulation modeling of the Martian atmosphere (Mischna & Shirley, 2017; Newman et al., 2019) indicates, for Mars at least, that the momentum equation should instead be written as follows:

$$\frac{DU}{Dt} = -2\,\omega_\alpha \times U - \frac{1}{\rho} \nabla p + g + cta + F_r \tag{2}$$

Here *cta* stands for "coupling term acceleration," as derived in (Shirley, 2017a):

$$cta = -c \int_{t=0}^{t=1\,s} \left(\frac{DL}{Dt} \times \omega_\alpha\right) \times r\, dt \tag{3}$$

In equation (3), *DL/DT* (hereinafter, *dL/dt*) represents the vector rate of change of the planetary orbital angular momentum (*L*) with respect to the *solar system barycenter* (SSB), which is the origin of the solar system inertial frame. As in equations (1) and (2) above, $\omega_\alpha$ represents the angular velocity of the planetary rotation, which for the Earth is a vector of magnitude 7.292115 x $10^{-5}$ rad/s directed along the axis of rotation. The position vector *r* of equation (3) identifies a particular location on, within, or above the surface of the rotating planet. It is referenced to a conventional body-fixed, rotating cartesian system, with origin at the center of the planet. The integration with respect to time yields units of m s$^{-2}$ (i.e., acceleration), which are efficient for numerical applications and are optimal when making comparisons with other forces acting. The leading parameter (*c*) is a scalar coupling efficiency coefficient whose value is to be determined by comparisons of numerical modeling outcomes with observations (Mischna & Shirley, 2017; Appendix 1).

The acceleration field described by equation (3) is projected upon a view of the Earth in Fig. 1. The cross product *dL/dt* × $\omega_\alpha$ lies within the equatorial plane of the rotating planet. The triple product *dL/dt* × $\omega_\alpha$ × *r* of equation (3) yields local acceleration (*cta*) vectors, here displayed for a sample of equally-spaced surface locations, each of which is specified by its unique Cartesian components (*r*).

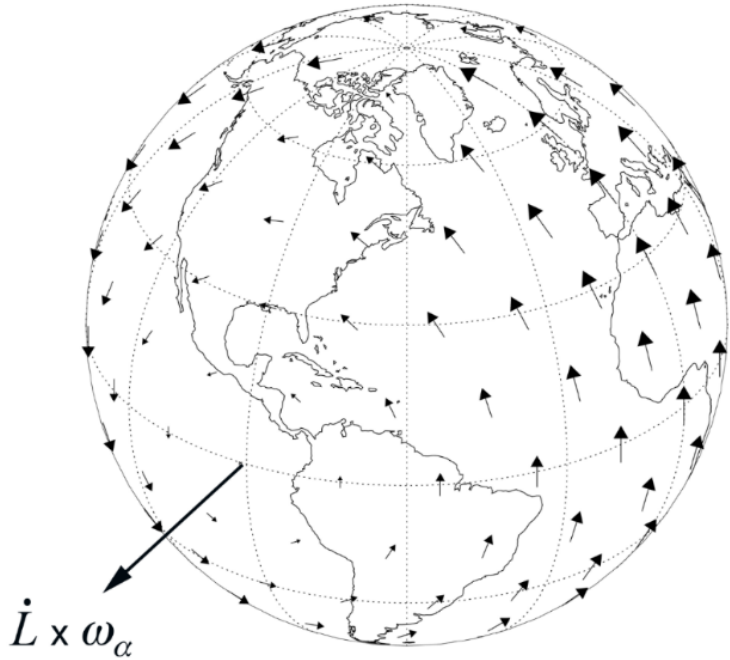

**Figure 1**. Instantaneous horizontal surface accelerations (*cta*), resulting from equation (3), as projected on a view of planet Earth. Viewed from this global perspective, the *cta* are seen to exert a torque about an axis coincident with the cross product $d\mathbf{L}/dt \times \boldsymbol{\omega}_a$, which lies within the equatorial plane. As the planet rotates, from west to east (i.e., from left to right), the direction of the acceleration at locations not on the equator will cycle in azimuth.

Methods for calculating the vector data displayed in Fig. 1 are described in Appendix A of Mischna & Shirley (2017). The orbital angular momentum and its time derivative are readily obtained from positions and velocities data generated by JPL's online Horizons ephemeris data system (https://ssd.jpl.nasa.gov/horizons). A series of coordinate transformations is required to represent the accelerations data in the terrestrial body-fixed (rotating) coordinate system for the date and time specified; the required transformations and sequence of operations are detailed in Appendix A of Mischna & Shirley (2017).

We now draw the reader's attention to several key aspects of the acceleration field of equation (3) and Fig. 1. Because the acceleration arises as a cross product with *r*, which points radially outward, the accelerations are everywhere directed tangentially to a spherical surface. Thus the *cta* have negligible vertical components of acceleration, and do not locally significantly modify the value of *g*. The *cta* in this way differ from the tidal forces (which typically include vertical components).

Let us initially suppose that the acceleration field of Fig. 1 is "fixed in space." We visualize the planet spinning beneath, or within, the vector field of Fig. 1. In consequence, over one (approximately) diurnal cycle, at any specific location not on the equator, the direction of the acceleration will cycle in azimuth. In the northern hemisphere, the acceleration is directed sequentially southward, westward, northward, and eastward. We will refer to this as the diurnal cycle, or the diurnal variability, of the *cta*. The cycle time most often approximates the sidereal day of 23.93447 hr, but this may vary significantly during torque reversals, as discussed below. Many other forms of temporal variability are superimposed upon this fundamental quasi-diurnal cycle. A number of these are described in Section 3 below.

The global pattern of the *cta* (Fig. 1) is reminiscent of the force diagram for a classical belt and pully system, with the axis of the pully coincident with the cross product *dL/dt* × $\omega_\alpha$. The *cta* are thus identified as comprising a torque acting about an axis lying within the equatorial plane. If the torque were to remain steady over long intervals of time, precessional motions of the spinning planet would be an expected consequence. However, the orbit-spin coupling torque is characterized by frequent reversals in direction. It is in this way unlike the well-known tidal torque acting continuously on the equatorial bulge of the Earth to produce the precession of the equinoxes.

Reversals in the direction of the torque come about when the dynamical forcing function *dL/dt* changes sign (Shirley & Mischna, 2017; Mischna & Shirley, 2017). With reference to Fig. 1, during such a transition, we must visualize the shrinkage and disappearance of the *cta* (surface acceleration) vectors, followed by their re-emergence with reversed directions. The times of disappearance and reversal of the *cta* have previously been characterized both as transitional intervals, and as a "relaxation phase," in which the active forcing by *cta* disappears for some interval of time.

We return to the topic of the time-variability of the torque in Section 3 below. More detailed discussions of the topics of this section may be found in Shirley (2017a); Mischna & Shirley (2017); Shirley & Mischna (2017); and Shirley, McKim, et al. (2020).

2.3. Quantitative aspects

The relative importance of the multiple forces that determine atmospheric motions under varying conditions may be estimated through *scaling analysis* (Holton, 1992). Such an analysis reveals that the pressure gradient force ($-\frac{1}{\rho} \nabla p$) and the Coriolis force ($-2\, \omega_\alpha \times U$) are typically the largest forces acting within the terrestrial atmosphere. Both attain magnitudes of ~ $10^{-3}$ (in units of acceleration per unit mass, i.e., in m s$^{-2}$) (Holton, 1992, Table 2.1). The approximate balance of these (the geostrophic balance) relates the atmospheric pressure field and the horizontal wind velocity in large-scale extratropical systems.

We now provide details of a calculation of the magnitude of the *cta* term of equation (3) using dynamical values (*dL/dt*) from the year 2020. As a first step, we obtain the instantaneous value of *dL/dt* for the Earth for 2020.857 (8 November 2020). (For convenience in the following calculations, and for ease in plotting, our unit mass is set equal to the Earth mass, i.e., 5.972 x $10^{24}$ kg).

Cartesian components of the *dL/dt* vector for the above date, represented in the Earth equatorial system, are then [3.7096 x $10^5$, -8.3207 x $10^5$, 1.0610 x $10^7$] in units of kg m$^2$ s$^{-2}$.

We next form the cross product of this vector with $\omega_a$, which in equatorial coordinates has the value [0, 0, 7.292 x $10^{-5}$] in units of rad/s. We obtain for $dL/dt \times \omega_a$, as illustrated in Fig. 1, component values [-401, 27, 0], with a resultant of magnitude ~403. Integration with time over a span of 1 s (as discussed above) effects a conversion to units of m $s^{-2}$ (acceleration) without altering the numerical values of the components.

Two further multiplications are needed to complete the calculation. The triple product of equation (3) requires us to obtain the cross product of $dL/dt \times \omega_a$ with $r$. If we choose a location on both the prime meridian and the equator (i.e., 90° removed from the longitude of emergence of the $dL/dt \times \omega_a$ vector in Fig. 1), then $r$ = [6.3781 x $10^6$, 0, 0], and the magnitude of the triple product becomes 2.564 x $10^9$.

Our final value for the acceleration on 8 November 2020 due to orbit-spin coupling is obtained by forming the product of the above very large acceleration with the coupling efficiency coefficient $c$ of equation (3). This coefficient is best found through an iterative comparison of modeling outcomes with observations (Mischna & Shirley, 2017, Section 4). However, no such comparisons have yet been performed in terrestrial applications. We thus consider that the current best available estimate for $c$ to be the value previously found in numerical investigations of the Mars atmosphere (Mischna & Shirley, 2017; Newman et al., 2019), which is $c$= 5.0 x $10^{-13}$. (Further discussion and justification of this choice may be found in Appendix 1). Completing the multiplication, we obtain a peak acceleration of 1.2 x $10^{-3}$ m $s^{-2}$ for the date in question. As we will show below in Section 5, the estimated average magnitude of the *cta* during 2020 is about 40% of this value, or ~5.1 x $10^{-4}$ m $s^{-2}$.

The above calculation reveals that the *cta* may occasionally be of the same order of magnitude as the Coriolis and pressure gradient forces in the Earth atmosphere. Accelerations of this magnitude clearly cannot be neglected.

A comparison of the acceleration calculated above with the peak acceleration of the lunar tide-raising forces is also of interest. The tidal acceleration produced by Earth's Moon at close approach (perigee) is $\leq$ 1.3 x $10^{-5}$ m $s^{-1}$, which is smaller by a factor of ~100 than the *cta* peak acceleration for 2020 indicated above. This difference will represent a factor of importance in our discussions of oceanic processes and effects in Section 6 below.

Additional discussion of the role and numerical value of the coefficient $c$ may be found in Appendix 1, which compares orbit-spin coupling torque amplitudes and atmospheric parameters at Mars with corresponding values for the Earth. The estimation of $c$ is discussed more comprehensively in Shirley (2017a), and in § 4 of Mischna & Shirley (2017).

2.4. The Dynamical System

To understand the sources of the orbit-spin coupling torques applied to the Earth system, it is first necessary to take account of various motions taking place within the larger dynamical system. A sketch of the dynamical system consisting of the Earth and Moon, the Sun, and the solar system barycenter (SSB) is provided in Fig. 2. Taking a top-down approach, we begin with a description of the solar system inertial frame, which is the fundamental coordinate system of Newtonian dynamics. The origin of the solar system inertial frame, or "local standard of rest," is coincident with the solar system barycenter. (For reference, the motion of the SSB about the gravitational center of the Milky Way galaxy corresponds to that of a point mass containing all of the mass of the solar system). Here we have sketched in a pair of coordinate axes for the inertial frame, in blue. This frame is without rotation, and is hence "fixed in space," with respect to distant objects in the universe.

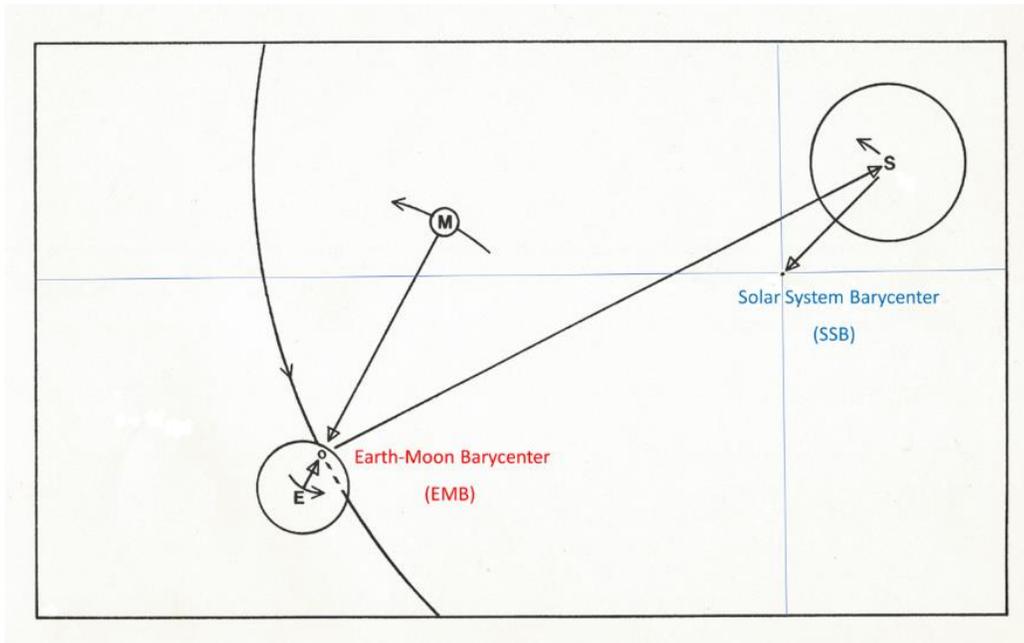

**Figure 2**. The dynamical system consisting of the Earth (E), the Moon (M), the Sun (S), and the solar system barycenter (SSB) (not to scale). The Earth and Moon orbit their common barycenter (EMB), while that barycenter orbits the Sun. The Sun in turn orbits the solar system barycenter. The first scientific description of the solar motion appeared in Book III of the Principia (Newton, 1687).

As indicated by a small curved arrow, the Sun is not at rest in this inertial system, but instead pursues a orbital trajectory about the "fixed" solar system barycenter. The somewhat irregular looping motion of the Sun about the SSB (Jose, 1965; Fairbridge & Shirley, 1987) is mainly due to the presence of the giant planets, whose orbital momenta together account for about 98% of the angular momentum of the solar system. The peak displacement of the Sun's center from the SSB is a little over 2 solar radii. This occurs when most or all of the giant planets are grouped to one side of the system. Interestingly, the Sun's orbital velocity is highest when it is most distant from the SSB, and lowest when it is closest to the barycenter. The close-approach events, which occur at intervals ranging from 15 to 25 yr, have been labeled "peribacs" (cf. Fairbridge & Shirley, 1987). Quite large quantities of orbital angular momentum (OAM) are gained, and lost, by the Sun, during these bidecadal-timescale orbital cycles (Table 1).

Table 1 lists the magnitudes of representative solar-system-dynamical quantities, in comparison to the magnitude of the axial angular momentum (AAM) of the terrestrial atmosphere (Lambeck, 1988). We highlight at this point the 14-orders-of-magnitude difference between the AAM and the angular momentum ($L$) of Earth's motion about the solar system barycenter, as the sheer size of the orbital reservoir provides additional perspective on the issue of the quantitative sufficiency of the orbit-spin coupling mechanism. Injection within the atmosphere of even a tiny fraction of the planetary OAM is likely to be of considerable geophysical significance. The torque of equation (3) and Fig. 1 effects such a transfer.

| Parameter | Angular Momentum (kg m$^2$ s$^{-1}$) |
|---|---|
| Axial Angular Momentum (AAM) of Earth's Atmosphere | ~1.2 x 10$^{26}$ |
| Earth's Orbit about the Earth-Moon Barycenter | 3.97 x 10$^{32}$ |
| Earth's Rotation | 5.86 x 10$^{33}$ |
| Moon's orbit about the Earth-Moon Barycenter | 2.8 x 10$^{34}$ |
| Earth's Orbit about the Solar System Barycenter | 2.63 - 2.69 x 10$^{40}$ |
| Sun's Orbit about the Solar System Barycenter | < 0 to >4.60 x 10$^{40}$ |
| Sun's Rotation | 1.92 x 10$^{41}$ |
| Orbit of Jupiter about the Sun | 1.9 x 10$^{43}$ |
| Solar System Total | 3.15 x 10$^{43}$ |

**Table 1**. Atmospheric, rotational, and orbital angular momenta. The atmospheric axial angular momentum (AAM) parameter quantifies the excess momentum (with respect to the underlying planetary surface) of westerly (zonal) flows in the atmosphere (Lambeck, 1988).

The orbital trajectory of the Earth-Moon center of mass, or barycenter, is likewise shown in Fig. 2. This barycenter lies along the line connecting the centers of the Earth and Moon. Its orbital motion about the Sun corresponds to that of a mass point containing all the mass of the Earth and Moon together.

The motions of the Earth and Moon about their common barycenter (EMB) are strongly perturbed by the gravitational influence of the Sun, and to a lesser but still significant extent by other planets such as Venus and Jupiter. Long and laborious calculations were needed, in earlier times, to calculate the position of the Moon with any accuracy. Isaac Newton is said to have told his friend Edmund Halley that "the problem made his head hurt, and he would think of it no longer" (Danby, 1997).

Interestingly, the OAM of the Moon, within the Earth-Moon system (EMS), is considerably larger than that of the Earth (Table 1). The Moon's OAM within the EMS exceeds in magnitude the angular momentum of Earth's rotation. By this measure, the Moon may be seen to exert considerable leverage within the Earth-Moon system.

The Earth and Moon are gravitationally bound to the Sun. In consequence, as the Sun orbits the solar system barycenter, the orbits (and trajectories) of the Earth, the Moon, and the EMB must necessarily follow along. With reference to Fig. 2, consider the situation a few years later, when the body of the Sun has shifted to the other side of the SSB. The trajectory of the EMB will likewise have shifted to the left, significantly increasing the distance between the EMB and the SSB (we reiterate that the SSB is by definition the origin of the coordinate system of this dynamical system). Because the numerical value of the OAM depends on distance, as well as on mass and velocity, this means that Earth's OAM with respect to the SSB (and inertial frames) is not constant, but instead varies significantly with time. Important aspects of this variability are detailed in Sections 3, 4, and 5.

## 3. Variability of the orbit-spin coupling torque on semi-diurnal to annual timescales

3.1 Orbit-spin coupling torques in 2020

Orbital angular momentum is continuously exchanged between the Sun and planets, due to mutual gravitational perturbations. As a result, the Earth's orbital angular momentum ***L*** and its rate of change *d**L**/dt* exhibits variability on all time scales. *d**L**/dt* for any solar system body may easily be calculated for intervals within 10,000 yr of the present, for any major solar system body, using source data of positions and velocities obtained from the JPL on-line Horizons ephemerides system (Giorgini et al., 1995). Before turning to a consideration of the cross product *d**L**/dt* × ***ω**_a*, of equation (3), yielding the *cta* and torque of Fig. 1, we must first briefly consider the *d**L**/dt* waveform itself, which has previously been characterized as the dynamical forcing function for the torque (Shirley & Mischna, 2017).

3.1.1. *d**L**/dt*

Figure 3a illustrates the terrestrial *d**L**/dt* waveform for the year 2020 (curve in red). The sign of the *d**L**/dt* waveform is positive at times when the Earth is gaining orbital angular momentum at the expense of other members of the solar system family. Conversely, the Earth is yielding up OAM at times when the sign of the *d**L**/dt* waveform is negative.

The *d**L**/dt* waveform of panel 3a exhibits amplitude modulation with periods of about 0.5-1 month and about half a year. The sources of both of these oscillations are easily understood with the aid of Fig. 2. The quasi-monthly oscillation of Earth's *d**L**/dt* waveform, shown in red in Fig. 3a, exhibits 12 positive peaks and 13 troughs. This oscillation is due to the presence of Earth's Moon. As the Earth and Moon orbit the EMB, their distances and orbital velocities (and hence their angular momenta) vary with respect to the SSB. Effects of the lunar synodic monthly cycle on the resultant torque will be discussed in greater detail in Section 3.1.4 below.

The longer term, quasi-semiannual modulation of Earth's *d**L**/dt* waveform in Fig. 3a is principally attributable to changes in distance between the Earth and the SSB. At the time illustrated in Fig. 2, the Earth (and the EMB) are closer to the SSB than to the Sun. Approximately 6 months later, when the Earth is on the other side of the Sun, the distance separating the Earth and the SSB will be considerably larger. Earth's orbital angular momentum (again, with respect to the SSB) will be correspondingly larger in magnitude at the latter time. The period of this cycle is a little longer than one year, due to the ongoing orbital motion of the Sun about the SSB (as indicated by the small arrow in Fig. 2). The Earth (and the EMB) will in consequence gain angular momentum, with respect to the SSB, for a period of a little over 6 months, as the Earth-SSB distance increases. The sign of the *d**L**/dt* waveform is positive in such cases. This factor largely accounts for the positive departures of the red curve seen in the latter half of 2020 in Fig. 3a.

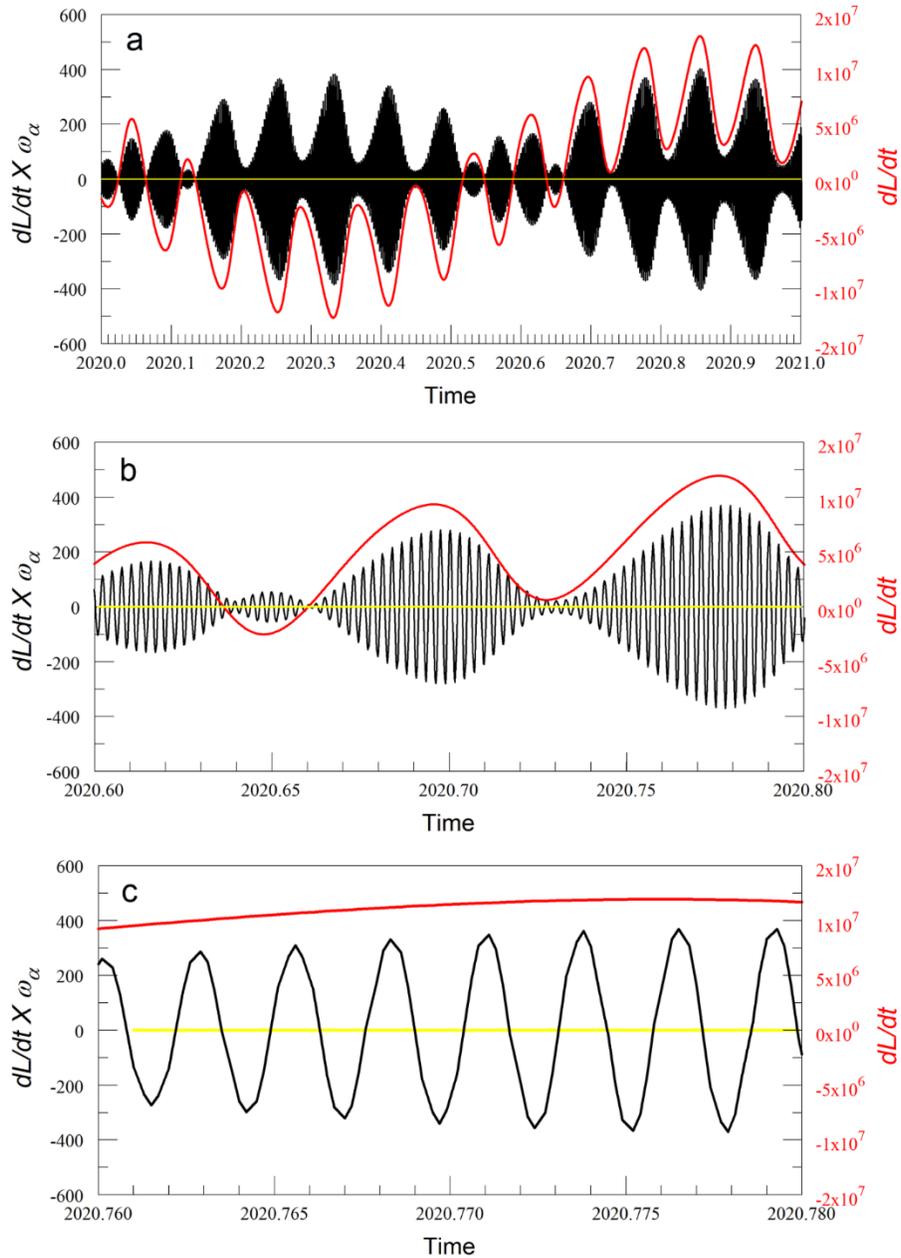

**Figure 3**. Time variability of terrestrial orbit-spin coupling torques in 2020. The waveform of the dynamical forcing function *dL/dt* is indicated in red, while the curve in black gives the x component of the cross product of *dL/dt* with the axial rotation angular velocity vector ($\omega_a$) (see discussion in text). Zoomed-in views of portions of the upper panel (a) are provided in panels (b) and (c). Reversals of the direction of the acceleration field of Fig. 1 occur at times when the *dL/dt* waveform intersects the zero line. Units are as presented in the calculations of the previous section: illustrated values of *dL/dt* have not been multiplied to include the numerical value of the Earth mass (5.97 x $10^{24}$ kg), and illustrated values of *dL/dt* × $\omega_a$ do not include the multiplicative factors of the Earth radius (6.371 x $10^6$ m) or the coupling coefficient *c*. The time step is 2 hr.

### 3.1.2. $d\mathbf{L}/dt \times \boldsymbol{\omega}_a$

The *dL/dt* waveform does not by itself provide a great deal of information about the diurnal and longer-term variability of the *cta* at any particular location on the planet. Recall from the discussion of Fig. 1 that the local acceleration cycles in azimuth over the diurnal period. We accordingly wish to resolve more directly the time-variability of the torque as expressed at a particular location on the Earth. To do so, in Fig. 3 we additionally plot (in black) the cross product $d\mathbf{L}/dt \times \boldsymbol{\omega}_a$, *as resolved in a standard coordinate system rotating with the planet*. We employ a time step of 2 hr, in order to resolve the variability at a particular location over the period of one day. As noted earlier, the cross product $d\mathbf{L}/dt \times \boldsymbol{\omega}_a$ is a vector lying in the equatorial plane, with no z component (Fig. 1). To illustrate the variability with time, it suffices to show only the x or the y component, as a function of time, as these are very similar, aside from their 90° shift in phase. We have chosen to show only the x component here.

The plotted positive extrema of the x component of the $d\mathbf{L}/dt \times \boldsymbol{\omega}_a$ vector plotted in Fig. 3 correspond to times of maximum northward acceleration along Earth's 90° E meridian of longitude. With the aid of Fig. 1, we may visualize the relationships as follows. We first imagine a clockwise rotation of the planet by 90°, leaving all else the same. The $d\mathbf{L}/dt \times \boldsymbol{\omega}_a$ vector will then emerge at longitude 0° and will have cartesian components [*r*, 0, 0] in the rotating coordinate frame. Under those conditions, the maximum northward accelerations will be found 90° to the east, i.e., above 90° E longitude.

The juxtaposition of the *dL/dt* (red) and $d\mathbf{L}/dt \times \boldsymbol{\omega}_a$ (black) curves in Fig. 3a clearly shows the relationship of the forcing function curve (*dL/dt*) to the amplitude variability of the local acceleration. We here distinguish 17 "pulses" of varying duration in 2020, each separated by brief intervals of much reduced torque amplitude.

The second and third panels of Fig. 3 are zoomed-in views of portions of the single-year torque amplitude history shown in the top panel (3a). The $d\mathbf{L}/dt \times \boldsymbol{\omega}_a$ waveform of panel 3a strongly suggests an interference (or beating) between two or more frequencies of oscillation, yielding the observed complex amplitude modulation of the torque as a function of time. Below we will show that this pattern results from the interference of torques arising 1) within, and 2) external to, the Earth-Moon system.

Panel (c) of Fig. 3 illustrates the variability of the x component of the cross product $d\mathbf{L}/dt \times \boldsymbol{\omega}_a$ over a period of about 8 days in October 2020. As previously noted, we can interpret the positive peaks of the curve as times of maximum northward acceleration at 90° E longitude. Noteworthy here is an increase in the amplitude of the peak acceleration, amounting to about 30%, taking place within the relatively short time interval shown.

### 3.1.3. Characteristic timescales and a conceptual analog model for atmospheric excitation

Earth's atmosphere possesses very little memory beyond about 1 month, as localized thermal anomalies and baroclinic disturbances normally decay significantly after the passage of this time interval (Wells, 2007). The oceans are thus considered to represent the principal source of memory for the coupled system. A characteristic timescale for the coupled ocean-atmosphere system may be ~5 yr (cf. Favorite & McLain, 1973; Hoffert et al., 1980; Wyrtki & Wenzel, 1984), although at the same time it should be noted that oceanic variability on significantly longer timescales is widely recognized (cf. Newman et al., 2016).

The orbit-spin coupling torque of Fig. 3 exhibits strong variability on timescales of about 0.5-1 month. Considered in juxtaposition with the known characteristic timescale for the atmosphere, this carries strong implications. If we suppose, for arguments sake, that each large pulsation or injection of momentum may excite or be accompanied by a perturbation of the wind and pressure fields, then we immediately recognize that these episodes are unlikely to grow and die out independently in time, as separate entities or events. Instead, we must visualize a superposition of the effects of multiple pulses. Motions within the coupled ocean-atmosphere system at any given time may thus represent the integrated effects of multiple superimposed pulses of momentum injection, superimposed on the underlying solar-driven circulations. Such a situation seems well suited to the excitation of capricious variability of weather systems.

As an aid to understanding, we suggest the following analog mechanism and process. Consider the mechanism of a clothes washing machine equipped with a vertical agitator that alternately turns, clockwise and counterclockwise, to impart turbulent motions to the water contained within it. A single pulse by the agitator (comprising a torque) will excite wave phenomena within the water that will die out over some short period of time. Under normal operation, the agitator supplies reversing impulses of torque, which in turn give rise to disordered and complicated motions of the water within the machine. Turbulent vortices may be excited as a component of these motions. We consider that the mechanical analog system of the vertical agitator washing machine usefully illustrates the likely effects of the application to the Earth system of the reversing orbit-spin coupling torques of Fig. 3.

Given this perspective, we immediately recognize the difficulty of identifying simple causal relationships, for instance between discrete weather events and individual torque pulses or episodes (as illustrated in Fig. 3). Achieving the goal of identifying such relationships would require, at a minimum, a driven atmospheric numerical model resolving the time history of the torque and its effects, past and future, together with a continuously updated data assimilation scheme. While this appears feasible, it is at the same time clear that a considerable investment of resources and time will be required before any such capability may be developed. These topics will be addressed in greater detail in § 8 below.

Questions of characteristic timescales and system response are less worrisome on decadal to bidecadal timescales, as the complex interactions likely arising on weather-related timescales may be effectively smoothed out. For this reason, in § 4 and § 5, our attention will focus more narrowly on the topic of decadal to bidecadal variability.

3.1.4. Interference effects

We return now to the question of the origins of the strong interference-like amplitude modulation of the torque, on quasi-monthly time scales, as highlighted our description of Fig. 3b above. We can separate the orbit-spin coupling torques applied to the Earth into two categories, each arising due to different sources. Referring once again to Fig. 2: We can isolate the orbital accelerations arising solely within the Earth-Moon system (EMS), and then consider these separately from those which are mainly attributable to the accelerations of the EMB. The latter are more strongly affected by the solar motion about the SSB.

In certain classical problems of dynamical astronomy, it may be computationally advantageous to avoid explicitly taking account of the complicated ongoing relative motions of the Earth and Moon. For instance, if we wish to study the motions of the planet Neptune, our calculations will be greatly simplified if we choose to represent the Earth and Moon as a single point mass located at (and moving with) the barycenter of the Earth-Moon system. Recognizing this, the designers of JPL's online Horizons ephemerides system allow the user to obtain positions and velocities of the Earth-Moon barycenter separately, as well as for the Earth and Moon individually, with respect to the conventional inertial coordinate system centered at the SSB. In order to isolate the accelerations arising principally within the Earth-Moon system, we merely subtract the EMB accelerations from the accelerations applied to the Earth due to all causes. Figure 4 illustrates curves of the *dL/dt* × $\omega_\alpha$ x *r* components (as in Fig. 3), for the EMS and EMB components separately, along with their resultant, in black, which corresponds to the amplitude-modulated curve shown earlier in Fig. 3b.

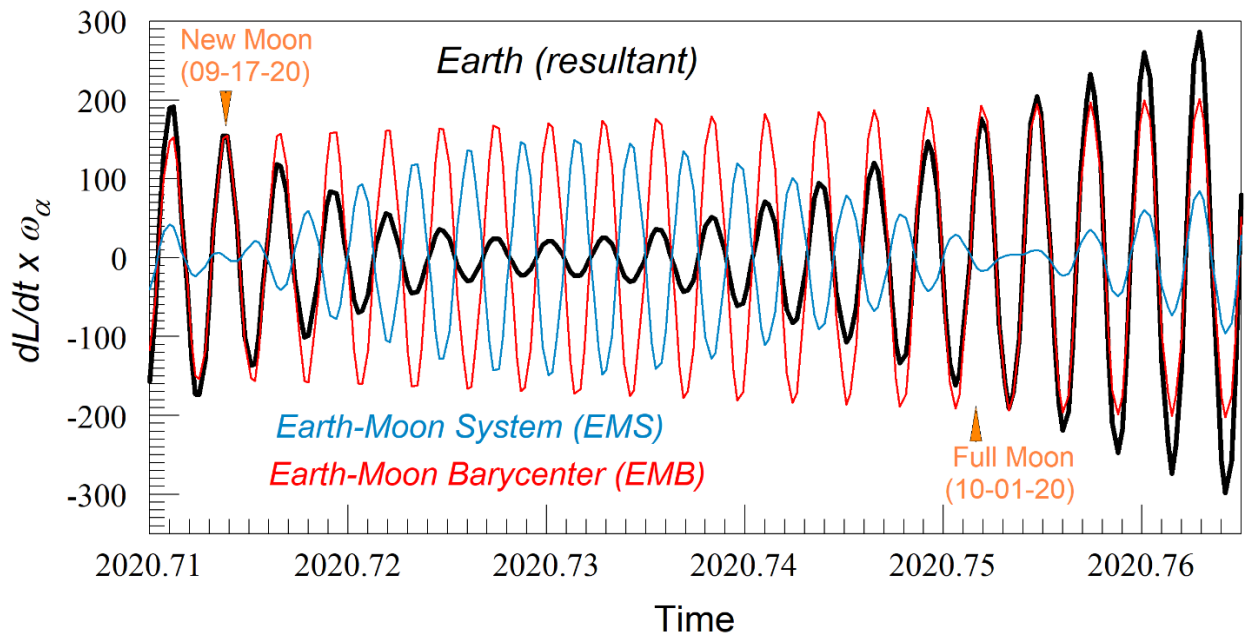

**Figure 4**. Superposition of the quasi-diurnal components of the two principal orbital acceleration sources gives rise to destructive interference, effecting an amplitude modulation of the orbit-spin coupling torque. Units are as in Fig. 3. The time span (0.055 yr) is about 20 days.

Included in Fig. 4 are a number of features worthy of note. The EMS component, in blue, is in phase with the EMB component (in red) both at the start of the interval, and at the end. In between, approximately between the times of the new and full Moon, the EMS component shifts rapidly to an out of phase condition. As indicated in Figs. 3a and 3b, this interval corresponds to a short period of time when the *dL/dt* waveform transitioned from positive to negative values, and back again. The phasing of the EMB component is unaltered during this episode. Thus the destructive interference that strongly reduces the amplitude of the resultant arises principally due to the EMS component.

3.1.5. Variability at fortnightly and synodic monthly periods: Comparisons with Observations

The times of the new and full Moon are also indicated in Fig. 4. These are times when the time variability of the EMS component of the *dL/dt* × $\boldsymbol{\omega_a}$ waveform falls to low values. The cycle time corresponds to about one half of the lunar synodic month of 29.5 days. This is traditionally known as the lunar fortnightly period, or cycle, of about two weeks duration. As is evident in Fig. 3, not every lunar fortnightly cycle results in the almost complete cancellation depicted in Fig. 4, which is associated with a significant reduction in the amplitude of the resultant torque.

Inspection of the EMS waveform for the full time span of this study (1860-2040) reveals that the fortnightly interference is a relatively prominent and robust feature of the time variability of the resultant torque. This carries important implications. While in this investigation we are primarily concerned with the variability of the torque on decadal to bidecadal timescales, it is appropriate to note here in passing some of the many past studies that have uncovered statistically significant fortnightly and/or lunar synodic monthly signals in atmospheric indices. Statistics of formation dates for hurricanes in the Atlantic (Bradley, 1964; Carpenter et al., 1972), typhoons in the Northwest Pacific (Carpenter et al., 1972), and depressions in the Indian Ocean (Visvanathan, 1966) bear systematic relationships to the lunar fortnightly cycle. Episodes of widespread heavy precipitation in the USA (Bradley et al. 1962) and in New Zealand (Adderly & Bowen, 1962) and South Africa (Visage, 1966) have been shown to preferentially occur at two opposed phases of the 29.5-d luni-solar synodic monthly cycle. Hanson et al. (1987) re-examined this question and found spatially variable phasing of the relationship across the USA. Records of precipitation from the Indian subcontinent reveal similar patterns (Berson & Deacon, 1965; Reddy, 1974; Roy, 2006), including a spatial dependence of phasing (Reddy, 1974; Roy, 2006). Also noteworthy are the results of Bigg (1963), who found a strong fortnightly cycle in measurements of atmospheric ice nucleus concentrations in the Southern hemisphere (in Australia, New Zealand, and South Africa). There are indications of a lunar phase signature in surface temperatures, both locally (Mills, 1966) and globally (Anyamba & Susskind, 2000). Other investigators document fortnightly cycle changes in indices characterizing the large-scale circulation of the atmosphere (Bryson, 1948; Mills, 1966; Miller, 1974).

It bears repeating that the lunar and solar tidal forces on the atmosphere are too small to satisfactorily account for the results cataloged above. Peak lunar tidal acceleration values reach only a little over $1 \times 10^{-5}$ m s$^{-2}$. Modern atmospheric GCMs therefore do not typically take account of gravitational tides. It likewise bears repeating that the *cta* and the orbit-spin coupling torques of Figs. 1, 3, and 4 bear no relationship to the tides, aside from certain similarities and coincidences in the time domain. As shown in the scaling analysis of Section 2.3, the *cta* are typically 1-2 orders of magnitude larger than the peak tidal accelerations.

In Section 4 we turn our attention to variability occurring on time scales from decades to centuries. However, before moving on, we must first make note of certain implications of Fig. 3 with respect to the topic of weather prediction. We focus here on the "subseasonal to seasonal prediction gap" ("S2S") (Mariotti et al., 2018). Present day numerical weather prediction techniques allow successful advance forecasting of weather events in middle latitudes for intervals of up to about 10 days in advance (Bauer et al., 2015; Alley et al., 2019). Supplementary empirical techniques (for instance, the observed phase states of atmospheric oscillations of longer periods, and the magnitude and distribution of sea surface temperature anomalies) allow insights and some forecasting capability with seasonal and longer lead times. The failure of current models to provide adequate forecasting capability beyond about 10 days in advance (i.e., within the subseasonal to seasonal prediction gap) has been attributed to a sensitive dependence on initial conditions within the atmospheric system (Lorentz, 1963). This effect is widely considered to impose a theoretical limit on weather prediction lead times (cf. Bauer et al., 2015; Scaife & Smith, 2018; Zhang & Kirtman, 2019; Alley et al., 2019). We suspect that the "subseasonal to seasonal prediction gap" may be more directly attributable to the failure of present-day numerical models and prediction schemes to include the accelerations due to orbit-spin coupling than to a sensitive dependence on initial conditions. Present-day prediction models take no account of the pulses of momentum added to atmospheric motions by the *cta* (Fig. 3). Our conjecture may be tested by new investigations incorporating the *cta* of equation (3) within existing numerical prediction models. We return to this topic in Section 8 below.

## 4. Variability of the torque on decadal to centennial timescales

Two dynamical sources of the of the resultant orbit-spin coupling torque applied to the Earth system may be separately resolved by means of the method described above in Section 3.1. As shown above in Fig. 4, and below in Fig. 5, the two components are identified with 1) motions of the Earth taking place due to gravitational interactions *within* the Earth-Moon System (EMS), as perturbed by the Sun and other bodies; and 2) motions, with respect to inertial frames, of the Earth-Moon barycenter (EMB). The time variability of the torques associated with motions of the EMB is largely driven by the ongoing motion of the Sun about the solar system barycenter, as previously illustrated in Fig. 2.

Figure 5 illustrates the relative contributions of the EMB and EMS contributions to the total torque (dark blue curve in panel 5c). The EMS contribution (Fig. 5b) is often the larger of the two, over extended periods of time; however, the peak values of the more variable EMB component can significantly exceed in magnitude those of the EMS.

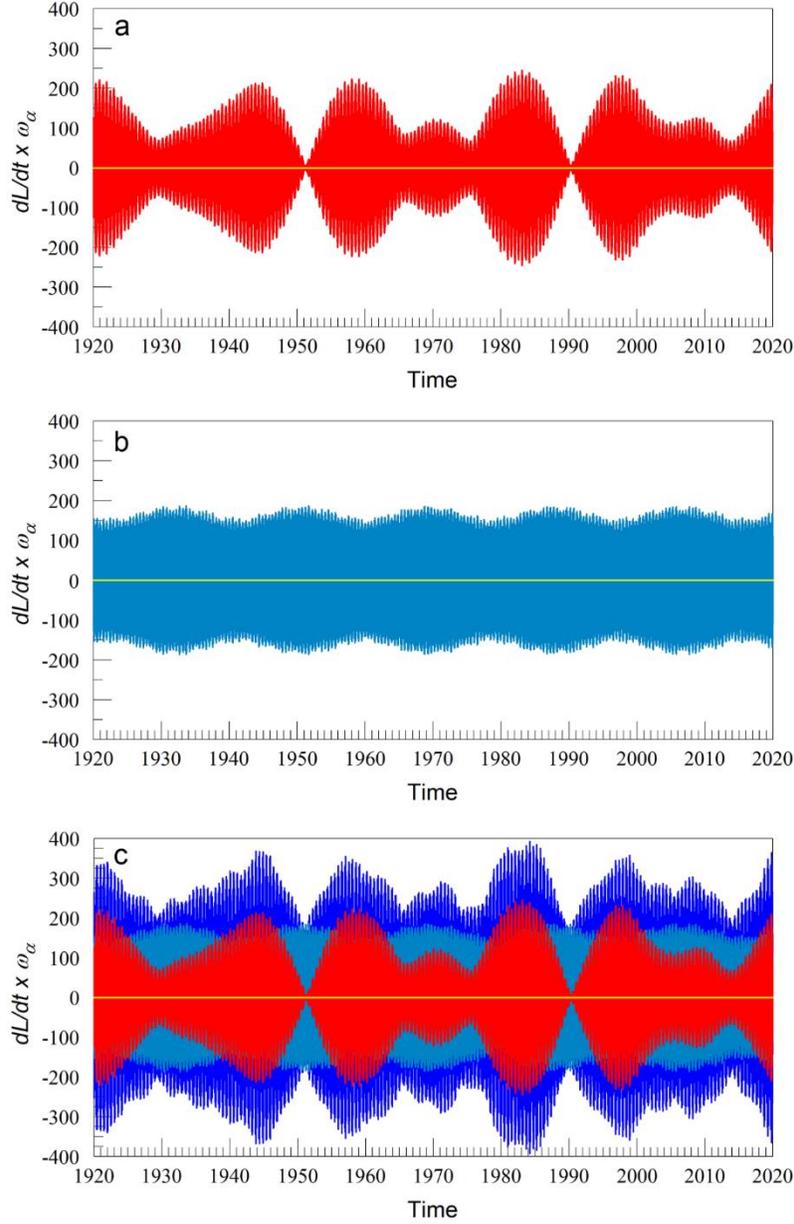

**Figure 5**. Separation of a century-long record of terrestrial orbit-spin coupling torques (1920-2020) into contributions due to the motion of the Earth-Moon barycenter (EMB) (panel a), and due to motions within the Earth-Moon system (EMS) (panel b). The dark blue curve of panel (c) illustrates the resultant torque, as shown earlier for shorter time intervals in Figs. 3 and 4, which is the sum of the two contributions. As in Fig. 3, we plot the x-component of the cross product $d\mathbf{L}/dt \times \boldsymbol{\omega}_\alpha$. Units are as in Figs. 3 and 4 above.

4.1. The Earth-Moon Barycenter (EMB) contribution

Considering first the EMB component of panel 5a, we immediately note the more complex variability with time, in comparison with the EMS component of panel 5b. The EMB component (Fig. 5a) nearly disappears in 1951 and around 1990. To understand the sources of this variability it is helpful to consider the trajectory of the Sun, with respect to the solar system barycenter, during these years, as illustrated in Fig. 6. The Sun's trajectory is largely determined by the orbital motions and relative positioning of the giant planets (Jose, 1965).

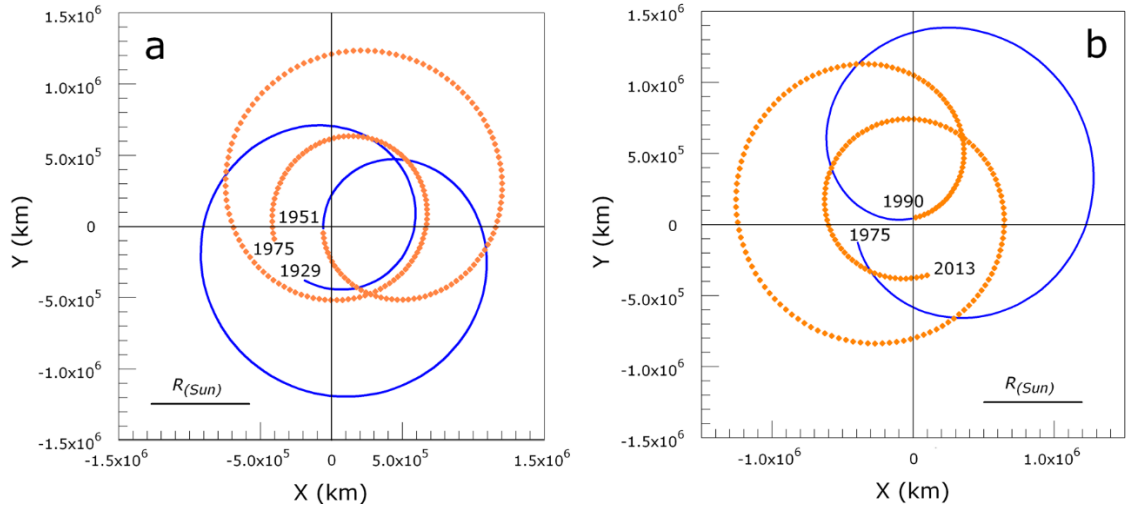

**Figure 6**. Ecliptic system polar plots of the orbital motion of the center of the Sun about the solar system barycenter (found at the origins of these plots, as in Fig. 2). a: 1929-1975. b: 1975-2013. Each plot includes two orbital cycles (Fairbridge & Shirley, 1987), which are color coded in orange and blue in each panel. The years of closest approach of the Sun's center to the solar system barycenter, termed "peribacs," are indicated. The Sun's radius (695700 km) is provided for scale.

In Fig. 6 we note immediately that the times in Fig. 5a when the EMB component nearly disappears (i.e., in 1951, and 1990) correspond in time to episodes of minimum separation of the Sun's center from the solar system barycenter. Other close approach episodes, in 1929, 1975, and 2013, show up as lesser minima of the torque waveform amplitude in Fig. 5a. The mean duration of the four orbital cycles shown is ~21 yr. Relationships purportedly linking the Sun's orbital motion with the excitation of the ~22-yr Hale magnetic cycle of solar activity were discovered and described by Jose (1965). Notable among many subsequent investigations are those of Fairbridge & Shirley (1987), and Shirley (2017b; see in particular Figs. 1 and 4 of that study).

Further discussion of the nature and effects of the EMB component of the orbit-spin coupling torque on the Earth system is deferred to Section 5 below. We turn now to a consideration of the EMS component as illustrated in Fig. 5b.

4.2. Contributions to the torque arising within the Earth-Moon System (EMS)

The most prominent feature of the EMS contribution of Fig. 5b is a regular oscillation of the waveform amplitude with a period of 18.6 yr. This modulation arises as a consequence of the retrograde revolution of the nodes of the Moon's orbit on Earth's equatorial plane, which is in turn a consequence of strong perturbations of the lunar orbital motion by the Sun. Figure 7 helps explain why this cycle is so strongly expressed in the $d\mathbf{L}/dt \times \boldsymbol{\omega}_a$ waveform of Fig. 5b.

Figure 7 shows the orientation of the lunar orbit plane, with respect to Earth's equator, at two times (T1 and T2) separated in time by 9.3 yr (one-half of the 18.6 yr period of revolution of the nodes). At time T1, the declination ($\delta$) attains its largest value (28.6°). The minimum value occurs ~9.3 years later, at time T2. Note in Fig. 7 that the angle between the rotation axis ($\omega_a$) and the vector normal to the orbit plane ($L$) is larger at time T1 than at time T2. From the properties of vector multiplication, we recall that the cross product of two vectors grows with the angular separation between them, reaching its largest magnitude when the separation angle is 90°. Accordingly, with all else being equal, the magnitude of the T1 orbit normal cross product with $\omega_a$ will be larger by a factor of ~1.5 than that found for time T2. This factor is largely responsible for the amplitude modulation of the EMS component $dL/dt \times \omega_a$ waveform of Fig. 5b. Dates of peak values of the lunar declination $\delta$ (within the timespan of Fig. 5b) include: 1932.02, 1950.63, 1967.24, 1987.86, and 2006.47.

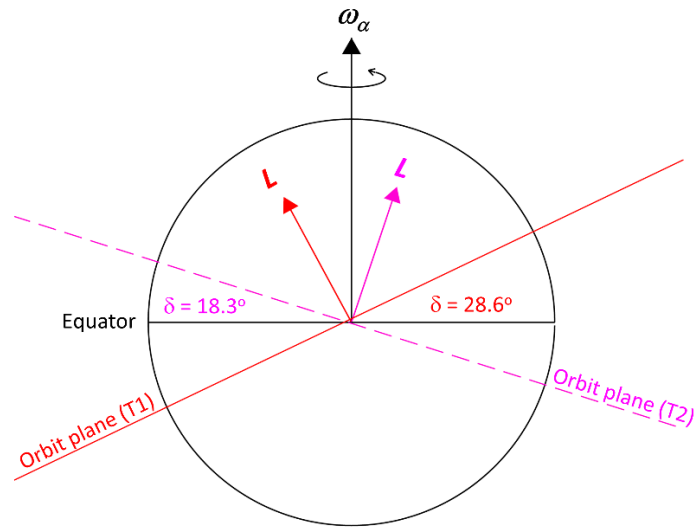

**Figure 7**. Changes of orbital geometry within the Earth-Moon system associated with the 18.6-yr lunar nodal cycle. The maximum declination ($\delta$) of the lunar orbit plane (shown in red) is achieved at Time 1 (T1). At such times, terrestrial observers will see the Moon reach its highest elevation in the night sky. At Time 2 (T2), indicated in purple, 9.3 years later, the declination angle is reduced (by ~10.3°). Arrows normal to the orbital planes are vector representations of the orbital angular momentum ($L$) of the lunar motion, with respect to the Earth's axial rotation angular velocity vector ($\omega_a$), at times T1 and T2 (see discussion in text). (Note: The vector angular momentum $L$ grows and shrinks with changes in the orbital velocity, while the instantaneous plane of the orbit remains nearly the same. The $z$ component of $L$ thus typically exhibits the largest changes of the 3 components. $dL/dt$ often thus lies nearly parallel to $L$. Thus, in Fig. 7, we have for simplicity plotted the orbital angular momentum vector $L$, rather than its more variable time derivative $dL/dt$).

The 18.6-yr $M_N$ component of tidal theory has the same cycle time as the EMS component of the orbit-spin coupling torque. The EMS variability (Fig. 5b) and the 18.6-yr tidal cycle both arise due to the same cause, as illustrated in Fig. 7. The shared cycle period may naturally lead to misunderstandings. Thus, at this point, we wish to highlight and emphasize key qualitative and quantitative differences between the $M_N$ tides and the EMS component of the *cta*.

$M_N$ is the 12$^{th}$ largest component in tidal theory. Its amplitude is about 4% of that of the lunar semi-diurnal tide ($M_2$). An excellent discussion is provided in (Ray, 2007). Its principal effect is to modulate the amplitude of the diurnal and semi-diurnal tidal waves, as measured in higher latitudes, as the tidal bulge due to the Moon shifts to a more poleward location during the high declination phase of the 18.6-yr cycle (i.e., at time T1 of Fig. 7). The amplitude of the $M_N$ component in the oceans has been measured using satellite altimeter data (Cherniawsky et al., 2010). It is quite small, being everywhere less than about 3.5 cm in amplitude. Ocean tidal currents, arising due to this cause, at such a low frequency, are expected to be vanishingly small (Ray, 2007). The $M_N$ tidal acceleration (estimated as a percentage of the $M_2$ peak value) is $\leq 1 \times 10^{-7}$ m s$^{-2}$. On the basis of a scaling analysis such as that of Section 2.3, one may reasonably conclude that this tiny acceleration is likely to be swamped by the effects of other, much larger, horizontal forces acting within the atmosphere. Atmospheric GCMs therefore typically neglect even the largest gravitational tides. To sum up: It is difficult to confidently identify a mechanism by which $M_N$, the 12$^{th}$ largest component of the tides, should give rise to detectable signals in atmospheric indices.

From a quantitative standpoint, the EMS component (Fig. 5b) of the total torque acting on the Earth system due to orbit-spin coupling is a much better candidate mechanism for the excitation of bidecadal variability at the 18.6-yr period. Figure 5 indicates that the EMS component accounts for ~40% of the peak amplitude of the resultant torque waveform. Forty percent of the best-estimate *cta* peak acceleration value for 2020 obtained in Section 2.2 comes to ~5 x 10$^{-4}$ m s$^{-2}$. Peak values of the lunar nodal cycle orbit-spin coupling acceleration are thus ~5000 times larger than the acceleration associated with the $M_N$ tidal component as estimated above.

4.2.1. The 18.6-yr Lunar Nodal Cycle (LNC) in Atmospheric Indices and Observations

Controversy has long attended claims of a link between atmospheric variability and the 18.6-yr lunar nodal cycle, principally due to the quantitative insufficiency of tidal explanations, as detailed above. Nonetheless there is an extensive literature on this topic. In this section, we provide a high-level overview of prior results from observations of Earth's atmosphere. Oceanic observations showing strong bidecadal variability at the lunar nodal period have likewise been controversial, but are now becoming more widely accepted. A number of key ocean science results are reviewed in Section 6 below.

The 18.6-yr lunar nodal cycle (LNC) has been detected in the motions, spatial distributions, and time variability of atmospheric pressure systems (Rawson, 1907, 1908, 1909; Bryson, 1948; Currie, 1982; O'Brien & Currie, 1993; Mazzarella & Palumbo, 1994; Cerveny & Shaffer, 2001; da Silva & Avissar, 2005; Yasuda, 2018).

The 18.6-yr LNC has also been detected in rainfall, precipitation, and flooding data, in multiple locations. An 18.6-yr cycle is prominent in monsoon rainfall in India (Campbell et al, 1983; Currie, 1984; Vines, 1986; Mitra & Dutta, 1992) and in precipitation in the USA (Vines, 1982; Currie & O'Brien, 1988, 1989, 1990a, 1990b, 1991, 1992). The phasing of the 18.6-yr LNC effects, with respect to location on the planet, has been found to be non-stationary. Regional differences in spatial phasing are found for monsoon rainfall in India (Mitra & Dutta, 1992), and for the LNC signal in rainfall in South Africa (Tyson et al., 1975; Currie, 1993; Malherbe et al., 2014). The LNC signal is also found in precipitation records from Australia (Currie & Vines, 1996), from China (Hameed et al., 1983; Currie & Fairbridge, 1985; Currie, 1995, 1996), from South America (Agosta, 2014), and from Europe (Vines, 1986) and the western Mediterranean (Mazzarella & Palumbo, 1994).

Many instrumental temperature records likewise exhibit statistically significant 18.6-yr LNC signals. Regional variations in phasing, as with precipitation, are frequently found; for instance, the LNC phase in temperatures east of the US Rocky Mountains differs from that to the west (Currie & O'Brien, 1992). Relationships to the LNC are reported in Keeling & Whorf (1997) for a global temperature anomaly record, and by Davis & Brewer (2011) in satellite measurements of temperatures. Regional temperature studies finding LNC signatures come from central Canada (Guiot, 1987), from areas surrounding the Gulf of Alaska (Wilson et al., 2007; McKinnell & Crawford, 2007), from the North Atlantic (Ynestad, 2006), from the western Mediterranean (Mazzarella & Palumbo, 1994), and from various regions in the USA (Bell 1981a, 1981b; Currie, 1981, 1993, 1996).

Other noteworthy studies have uncovered relationships linking the 18.6-yr LNC with North Atlantic tropical cyclone frequency (Currie, 1996), and with the flows of the Nile and St. Lawrence rivers (Hameed, 1984; Currie, 1994; Kondrashov et al., 2005). Last but not least, the 18.6-yr LNC cycle is also detected in studies of tree ring variability from multiple locations. An influential study of tree rings and the drought area index (DAI) for the western USA by Cook, Meko, & Stockton (1997) confirms the results of earlier investigations by Currie (1981, 1984a, 1984b, 1996), by Bell (1981a, 1981b), and by Stockton et al. (1983). The LNC is likewise found in tree ring records from Patagonia (Currie, 1983), from Argentina and Chile (Currie, 1991a), from areas surrounding the Gulf of Alaska (Wilson et al. 2007), from Mongolia (Davi et al., 2006), from Europe (Currie, 1992), and from Tasmania, New Zealand, and South Africa (Currie, 1991b).

Orbit-spin coupling for the first time provides a quantitatively plausible working hypothesis for the excitation of spatially widespread atmospheric bidecadal variability at the 18.6-yr periodicity of the retrograde revolution of the nodes.

## 5. Torque Magnitude and Multiyear Drought Episodes in the Western USA since 1860

Many subtle variations of the torque magnitude as a function of time may be discerned, with some difficulty, in Fig. 5c. In order to more clearly show the variability of the torque magnitude with time, we plot the resultant of the x and y components of $d\mathbf{L}/dt \times \boldsymbol{\omega}_\alpha$ in Fig. 8. Units are as in Figs. 3-5. Figure 8 extends the range of temporal coverage of Fig. 5 by 80 years, with the plotted data beginning in 1860 and extending to 2040. Two additional data types are plotted along with the torque magnitude in Fig. 8. The upward pointing arrows indicate past and future times of close approach by the Sun to the solar system barycenter. Close-approach epochs (peribacs) for the interval from 1920-2020 were shown previously in Figure 6. The deep torque minima of Fig. 8 correspond in time to the minima of the EMB component considered separately, as illustrated earlier in Fig. 5a. Dynamical causes for the association between peribacs and the EMB component torque minima were outlined above in Section 4.1. The correlation between solar peribacs (yellow arrows) and torque minima of Fig. 8 is therefore unsurprising and expected.

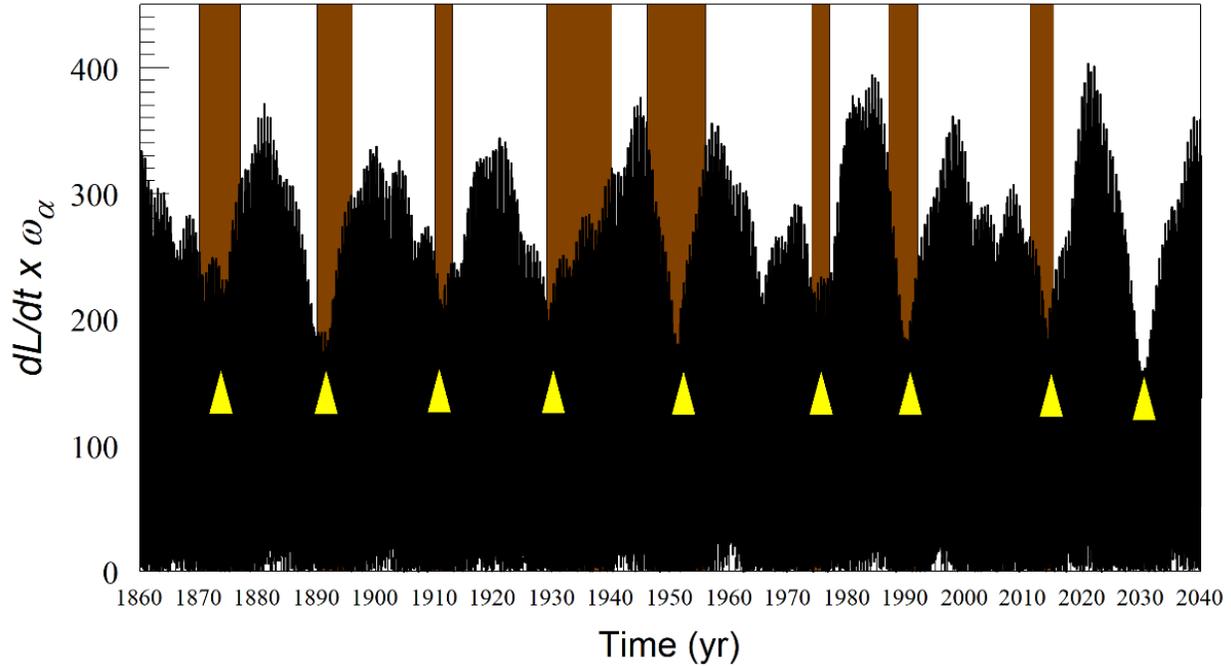

**Figure 8**. Earth system torque magnitude (1860-2040). Arrowheads (in yellow) indicate times of close approach of the Sun to the solar system barycenter (as shown earlier in Fig. 6). The next such close approach will occur in 2030. Vertical bars (brown) lie between the starting and ending times of the selected multiyear Western USA drought episodes of Table 2. Units are as in Figs. 3-5.

Vertical bars in Fig. 8 illustrate the timing and duration of the multiyear drought episodes of the Western USA listed in Table 2. The correspondence of Earth system orbit-spin coupling torque minima and multiyear episodes of widespread drought in the Western USA revealed in Fig. 8 has not previously been recognized and is the subject matter of the balance of this Section.

Every instance of a close approach by the Sun to the solar system barycenter since 1860 has been accompanied by an extended period of drought in the Western USA. However, not every instance of widespread drought in the Western USA can be associated with deep torque minima as illustrated in Fig. 8. At least three other multiyear episodes of widespread Western USA drought took place during the time period illustrated. One of these was already underway in 1860, extending from 1856-1865 (Fye et al., 2003; Herweijer et al., 2006). Another occurred in the years from 1897-1904 (Fye et al., 2003). A third multiyear episode, sometimes termed the "turn of the century drought" (Cook et al., 2010), extended from 1998 or 2000 until 2005 or later, with the quoted times depending on the parameters and definitions employed (cf. Martin et al., 2020). None of these occurred near the time of a close approach by the Sun to the SSB.

We recognize, with many prior investigators, that multiple causes are involved in the occurrence of multiyear drought episodes. Teleconnections, notably with equatorial Pacific sea surface temperatures (Namias, 1978, 1979, 1983; Trenberth et al., 1980; Cook et al., 2007) are often recognized and described. Even within the restricted context of the orbit-spin coupling mechanism under discussion, in addition to the 22-yr rhythm, statistically significant relationships linking drought occurrence with the 18.6-yr lunar nodal cycle (Section 4.2) are reported in multiple investigations (Currie, 1981, 1984a, 1984b; Bell, 1981a, 1981b; Stockton et al., 1983; Cook et al., 1997). The simple relationship revealed in Fig. 8 thus represents only part of the full story. In the following we review many results from prior studies in order to better understand and more fully characterize pertinent system behaviors. We begin with a reconsideration of the 22-yr rhythm of drought in the Western USA described in Mitchell et al. (1979) and in Cook et al. (1997).

| Drought Episode | Peribac Year | Duration (yr) | References |
|---|---|---|---|
| 1870-1877 | 1873.6 | 8 | Herweijger et al., 2006 |
| 1890-1896 | 1891.2 | 7 | Herweijger et al., 2006; Cook et al., 2007 |
| 1910-1912 | 1910.3 | 3 | Day, 1911; Borchert, 1971; Burnett & Stahle, 2013 |
| 1929-1940 | 1929.7 | 12 | Fye et al. 2003; Cook et al. 2007 |
| 1946-1956 | 1951.4 | 11 | Fye et al. 2003; Cook et al. 2007 |
| 1975-1977 | 1975.2 | 3 | Ratcliffe, 1977; Seager et al, 2015 |
| 1987-1992 | 1990.3 | 6 | Swain et al., 2014; Trenberth et al., 1988 |
| 2011-2015 | 2013.8 | 4 | Seager et al., 2015; Swain, 2015 |

**Table 2**. Multiyear episodes of widespread Western USA drought associated with peribacs (1860-2020). Dates and durations are from the cited references. The table does not include all of the multiyear drought episodes of the historic record since 1860 (see discussion in text). Additional considerations pertinent to the selection of the time interval for the present investigation are summarized in Appendix 2.

The 22-yr rhythm of drought in the western USA (Mitchell et al., 1979; Stockton et al., 1983; Cook et al., 1997) and its apparent association with the 22-yr Hale (double) sunspot cycle has puzzled investigators for many years, since (as previously noted) no viable candidate physical mechanism has yet been advanced to account for such a relationship. With respect to this issue, we briefly make note of the following. The epochs of Hale sunspot cycle minima employed in Mitchell et al. (1979) for their comparisons are listed in the appendix to that paper. The last 5 dates in their table are 1889.4, 1913.2, 1933.4, 1954.1, and 1976.8. These dates closely correspond to 5 of the peribac dates listed in Table 2 (1891.2, 1910.3, 1929.7, 1951.4, and 1975.2); the signed mean difference of the two series is only about +1.8 yr. Further, a similar relationship holds near the start of the period (1700 CE) analyzed by Mitchell et al. (1979) (see Fig. 1 of Shirley, 2017b), such that more than half of the Hale cycle epochs of Mitchell et al. (1979) correspond more or less closely to peribac dates. While it is beyond the scope of this paper to pursue this question, we favor an interpretation in which both phenomena (i.e., Hale sunspot cycle minima, and multiyear episodes of drought in the western USA associated with peribacs) are dynamically synchronized by a common mechanism (orbit-spin coupling), simultaneously operating on both subject bodies independently (Shirley, 2017b).

5.1. Drought observations and the causes of drought

"The central problem is how the high pressure areas are forced to recur persistently over the drought prone area." (Namias, 1979).

Persistent ridging gives rise to subsidence above affected areas, producing low rainfall, high temperatures, and stable atmospheric conditions.

Namias (1978, 1979, 1983) recognized that important episodes of severe drought were not primarily driven by local conditions, but are instead linked with global-scale variations of the atmospheric circulation that persist over extended periods of time. The observable phenomena have been characterized and labeled in different ways; Namias (1978) cited "runs of recurrent circulations," while Trenberth et al. (1980), referring to the persistence of large scale Rossby wave patterns, identified "stationary wave trains." Cook et al. (2007) preferred "stationary Rossby waves," while others focus on "strong ridging" (Cook et al. 2014), or "persistent ridging" (Swain et al. 2014). Considerable work has focused on anomalies of the circulation in the upper troposphere, or in the lower troposphere, or both. The phenomenon of the unusual persistence of high pressures adjacent to western North America during 2011-2015 was given a colorful label, becoming known as the "ridiculously resistant ridge" (Swain, 2015).

The persistence of the anomalous large-scale circulation patterns on seasonal and longer timescales strongly implies that some component of the coupled system (other than the atmosphere) must either 1) supply some form of system memory, or 2) constrain the system variability in some other way. In light of the large heat capacity of the oceans, much attention has been paid to a possible influence of sea surface temperatures on the large-scale circulation. Land-atmosphere coupling also may play a role (Namias, 1983; Basara et al., 2013); in this mechanism, soil moisture variations play a role in the maintenance of the circulations that promote drought (Oglesby & Erickson, 1989; Roundy et al., 2013; Zeng et al., 2019).

The record of past investigations of correlations of sea surface temperatures (SSTs) with drought is instructive. Earlier studies of individual droughts or small samples of droughts often found correlations with anomalous SSTs (Namias, 1978, 1983; Trenberth et al., 1988; Palmer and Branković, 1989). Subsequent investigations employing observed sea surface temperatures as specified boundary conditions for numerical modelling using GCMs were likewise often successful in reproducing drought conditions (Hoerling and Kumar, 2003; Herweijer et al., 2006; Cook et al., 2007). These simulations tended to corroborate previously suspected relationships between anomalous SSTs and the large-scale circulation on hemispheric scales. The demonstrated planetary-scale teleconnections emphasize that drought in the Western USA cannot be considered to be solely a regional phenomenon.

More recent simulations call into question the universality of proposed causal relationships between SSTs and drought occurrence. Seager et al. (2015) investigated the causes of the 2011-2014 California drought but could find no obvious SST forcing. Cook et al. (2014) likewise found only small contributions to the genesis of the dust bowl drought of 1934 from SSTs. Parsons & Coats (2019) describe detailed interactions of the ocean and atmosphere in observations and model data, finding complex relationships of droughts to El Nino occurrence and SSTs. Baek et al. (2017) found relationships linking droughts with multiple large scale atmospheric oscillations, including the PDO (Pacific Decadal Oscillation) and NAO (North Atlantic Oscillation), along with El Nino.

Thus, while correlations of drought occurrence with anomalous SSTs are too frequent to be dismissed, it is now evident that SSTs alone cannot solely account for drought occurrence; droughts apparently arise due to multiple causes. We return to the question of oceanic contributions in Section 6.

The foregoing brief discussion cannot adequately capture the full scope and range of the prior literature on drought occurrence. Many contributions and topics are not mentioned, and our discussion lacks nuance and detail. Each of the droughts of Table 2 were unique, for instance, in terms of areas affected and duration (Herweijer et al., 2006; Cook et al., 2007). The above review is included to provide context for the material of the next section, in which we present a working hypothesis for multiyear drought occurrence.

5.2 Working hypothesis for multiyear Western USA droughts coincident with peribacs

The one-to-one correspondence of drought occurrence with orbit-spin coupling torque minima of Fig. 8 suggests that multiyear droughts cannot be explained as a result of a "powering up" of the large-scale circulation. Instead, we must think in terms of a *diminution* of active forcing. Torque minima have been characterized as relaxation intervals, during which the excess momentum deposited within the circulation by the active forcing may be transferred from the atmosphere to the underlying planetary surface by frictional effects (cf. Shirley, McKim et al. 2020, Section 3.1.3), thereby reducing relative atmospheric angular momentum and energy. The predicted relaxation (Shirley, 2017a) of Martian global wind systems has been reproduced in numerical simulations (Mischna & Shirley, 2017). Low-torque periods of circulatory relaxation correspond to less-stormy conditions on Mars.

The proposed external forcing of terrestrial global wind systems by orbit-spin coupling is essentially continuous in time (Figs. 3, 5c and 8). The capricious normal variability of terrestrial weather and climate, with its great diversity of circulation patterns, may thus be understood, under the orbit-spin coupling hypothesis, to result, at least in part, from a continuous (and continuously variable) forcing by the external torque (Section 3.1.3). We then ask: How might the terrestrial atmospheric circulation respond to deep multiyear torque minima, as illustrated in Fig. 8?

We submit that a *reduction* in the *variability with time* of the large-scale circulation of the atmosphere may reasonably be expected, under conditions of significantly reduced forcing by the torque. A significant reduction in the variability with time of atmospheric flows (including meridional flows) may in turn plausibly give rise to reduced time variability in the longitudinal patterns and wavenumbers of atmospheric Rossby waves. Runs of recurrent circulations (Namias, 1983), and stationary wave trains (Trenberth et al., 1988) have both been associated with the strong subsidence and stable atmospheric conditions associated with multiyear drought episodes.

Under the above working hypothesis, we further suspect that surface effects and characteristics, including SSTs, land-atmosphere coupling mechanisms, and global topography, could exert a greater than normal influence on the morphology of the large-scale circulation, during deep torque minima, thereby helping to maintain the anomalously persistent anticyclonic circulations responsible for severe droughts.

In connection with statistical metrics of the large-scale circulation during drought episodes, the term "persistent anomaly" may be taken to be nearly synonymous with "a period exhibiting reduced time variability." While by no means conclusive, the following statistical comparison may yet be of interest. In describing geopotential height anomalies of the California drought of 2011-2015, Swain et al. (2014) note that "a vast geographic region centered in the Gulf of Alaska experienced 500-mb GPH anomalies that exceeded all previous values," and also that "standard deviation of the daily 500-mb GPH field was extremely low over much of the northeastern Pacific". Their observations prompt us to consider the magnitude and time variability of the torque during this time period, which may be characterized using similar statistics.

We first consider the 4-yr period 2011-2014, comprising the California drought years, as illustrated in Fig. 8, where the data is provided in 2-hr time increments. We find that the mean value of torque over this interval was 95.4 (in the units of Fig. 8), with a standard deviation σ of 62.0. The largest torque value within the interval was 262.1. These values may be compared with similar statistics for a high-torque interval; we chose the 4-yr interval from 2020-2023. In that case (see Fig. 8), the mean and standard deviation are 163.6 and 106.6, while the maximum value was 402.7. This comparison indicates that the *variability* of the torque, as measured by the standard deviation, was smaller by ~42% during 2011-2014 than during 2020-2023. Under the present working hypothesis, this is an expected outcome; reduced (and less variable) pulsed external input of momentum to the atmospheric system should be accompanied by reduced time variability of the large-scale circulation.

The results presented above in Figs. 3, 4, and 8 thus carry the following important implication: *The time variability of the large-scale circulation of the atmosphere may be driven by the time variability of the torque.*

The above working hypothesis may be validated or disqualified through targeted experiments employing numerical simulation in AGCMs and AOGCMS. We return to this topic in Section 8 below.

The orbit-spin coupling torque is highly deterministic. Incorporating this forcing within atmospheric and oceanic GCMs and coupled models should to some extent reduce the need to single out memory components within the Earth system that could be uniquely and directly responsible for the occurrence of multiyear drought episodes. Even so, it is clear that the heat capacity of the oceans must play some role. We now turn our attention to the topics of ocean-atmosphere coupling and bidecadal variability.

## 6. Orbit-spin coupling torques and coupled ocean-atmosphere dynamics

A number of relationships have been found linking Western USA drought occurrence with well-known oscillation modes of Earth's coupled ocean-atmosphere system, notably including ENSO (El Nino-Southern Oscillation) and the PDO (Pacific Decadal Oscillation). The negative phase of each of these oscillations is characterized by cold sea surface temperatures (SSTs) in the equatorial central and eastern Pacific. The physical origins of the observed relationships have long been obscure. ENSO and the PDO have been active areas of investigation for many years; for this reason, it is beyond the scope of this paper to attempt to comprehensively survey or summarize the entire field. Nonetheless, a consideration of the relationships of these modes to drought occurrence, in juxtaposition with orbit-spin coupling predictions and effects, may still be of interest. One goal here is to highlight certain connections that may potentially be explored in future investigations.

6.1. Basic considerations of air-sea interactions

The higher heat capacity of ocean waters, together with relatively slowly varying distributions of surface and subsurface temperatures and temperature anomalies, helps provide long-term memory for the coupled ocean-atmosphere system. The circulation of the oceans accounts for roughly half of the total equator-to-pole heat transport of the coupled system. This is principally accomplished by and within western boundary currents such as the Gulf Stream. Strong atmospheric baroclinic activity is observed above regions of strong gradients of SSTs (Nakamura et al., 2004; Frankignoul et al., 2011). Ocean temperatures are in this way capable of modifying atmospheric storm tracks (Sung et al., 2014).

The ocean circulation is not principally driven by differences of temperature or salinity (Munk & Wunsch, 1998); instead, it is strongly driven mechanically, by processes which are not yet fully understood. Surface currents are largely driven by winds (Munk, 1950). Observed large internal currents are thought to be driven principally by lunisolar tidal effects; these give rise to strongly turbulent motions and internal mixing (Ferrari & Wunsch, 2009), particularly when the currents interact with bathymetric topography. Internal diapycnal mixing due to this process is an active area of research (MacKinnon et al., 2017), as this mixing is energetically required to sustain observed oceanic meridional overturning circulations (Munk & Wunsch, 1998). Vertical mixing within ocean water masses alters both SSTs and nutrient availability, and is thus of economic as well as climatological significance. While tidal processes are much more important for the ocean circulation than for the atmosphere, there is one important point of correspondence: Just as detailed above in Sections 1 and 4.2 for the atmosphere, the very small tidal accelerations within the oceans associated with the 18.6-yr lunar nodal cycle are unlikely to excite significant bidecadal variability in oceanic currents or other indices (Ray, 2007).

6.2. Oceanic observations and the 18.6-yr lunar nodal cycle

Orbit-spin coupling torques are applied to the oceans directly (Shirley, 2017a). The *cta* may additionally give rise to indirect contributions to the oceanic circulation, by modulating atmospheric wind systems (see Sections 3.1.5 and 4.2.1 above). In this subsection we review prior observational results pertaining to the Earth-Moon system (EMS)-associated 18.6-yr LNC component only. To our knowledge, the variable cycle time of the EMB component has prevented any prior identification of oceanic bidecadal variability due to that source.

The 18.6-yr LNC is strongly expressed in North Pacific sea surface temperatures and air temperatures (Loder & Garrett, 1978; Royer, 1993; McKinnell & Crawford, 2007). Vertical mixing variations near the Kurile Straits and adjacent areas give rise to SST changes and salinity changes that can be conveyed eastward by the Kuroshio and Oyashio current systems, thereafter generating or impacting SST anomalies over much of the North Pacific (Osafune & Yasuda, 2006; Yasuda et al., 2006). Numerical modeling with OGCMs with parameterized mixing confirms and illuminates large-scale downstream effects (Hausumi et al., 2008; Tanaka et al., 2012; Osafune & Yasuda, 2012, 2013; Osafune et al., 2014, 2020). Strong mixing variability at the LNC period is also found for the Bering Sea (Foreman et al., 2006; Osafune & Yasuda, 2010) and Okhotsk Sea (Rogachev & Shlyk, 2018). Along the western coast of the USA, the LNC appear as a significant feature in SSTs measured at the Scripps pier (California) and in the California current (McKinnel & Crawford, 2007).

Loder and Garrett (1978) additionally show signatures of the LNC in SSTs along the Atlantic coast of the USA. The LNC also appears prominently in Arctic climate, i.e., in sea levels, in temperatures, and in sea ice extent, among other parameters (Ynestad, 2006; Ynestad et al., 2008), and in arctic ocean biomass (Ynestad, 2009). It is found in sea levels and SST in the Mediterranean (Mazzarella & Palumbo, 1994). Other LNC correlations at larger spatial scales are noted below.

6.3. ENSO and drought occurrence

ENSO (El Niño-Southern Oscillation) is the dominant mode of year-to-year climate variability on Earth (Horel & Wallace, 1981; McPhaden et al., 2006). The atmospheric component (the Southern Oscillation) involves a quasi-periodic see-sawing oscillation of atmospheric pressures in equatorial regions and in the southern hemisphere, where the eastern and western centers correspond approximately to Tahiti and to Darwin, Australia. El Niño is the name given to the positive phase of an alternating cycle of sea surface temperatures in the eastern Pacific (Rasmusson & Carpenter, 1982). The cold phase is termed La Niña. ENSO impacts weather and climate over much of the planet through shifting pressure patterns and storm track positions. Precipitation in the Western USA typically increases during the positive phase of El Niño, and decreases during the colder eastern tropical Pacific phase (La Niña) (Redmond & Koch, 1991).

A possible association of Western USA drought episodes with the cold phase (La Niña) of ENSO has been under investigation for many years, with most studies concluding that a connection exists (cf. Rajagopalan et al., 2000; Cole et al., 2002; Cook et al., 2004, 2007; Woodhouse et al., 2009). In a recent study, Parsons & Coats (2019) investigated the association between ENSO phenomena and the developmental evolution of multiyear droughts of the Southwestern USA, finding that El Niño events tend to both precede, and follow, multiyear droughts. A significant relationship between cool tropical Pacific SSTs and drought years was found.

Parson & Coats (2019), noting that exceptions occur, at the same time challenged the cold eastern Pacific SSTs paradigm, noting that the strong El Niño of 2015-2016 was not accompanied by anomalous precipitation in California or the Western USA, while the following non-El Niño winter was. Teng & Branstator (2017; see also Seager et al., 2015) likewise indicate that not all California droughts may be linked to negative phase ENSO conditions.

6.4. The Pacific Decadal Oscillation (PDO) and drought occurrence

The PDO is the first empirical orthogonal function (EOF) of the anomalies of monthly mean SST poleward of 20° N in the Pacific Ocean (Mantua et al., 1997; Schneider & Cornuelle, 2005; Newman et al., 2016). The PDO varies on interannual and decadal time scales. It is widely used to characterize decadal variability of Northern Hemisphere climate and the North Pacific ecosystem. The PDO takes the form of a horseshoe-shaped distribution of sea surface temperatures spanning the Pacific basin, with the ends of the horseshoe extending both northwest and southwest of the area dominated by the confluence of the Kuroshio and Oyashio current systems, east of Japan, in the western and central Pacific. As with ENSO, the negative phase of the PDO is associated with colder ocean SSTs in the eastern Pacific (adjacent to the North American continent). The colder ocean temperatures in the eastern Pacific are associated with atmospheric ridging over the Western USA. For this reason, the PDO, like ENSO, is often linked with drought occurrence in the Western USA (Cole et al., 2002; Goodrich, 2007; Wise, 2010; Nelson et al., 2011; Sung et al., 2014).

The PDO is recognized as a superposition of multiple physical processes acting on different timescales (Schneider & Cornuelle, 2005; Newman et al., 2016; Wills et al., 2019). Among these processes are forcing by zonal advection of temperature anomalies in the Kuroshio-Oyashio Extension on decadal timescales, forcing by variability of the Aleutian center of action associated with El Nino on ENSO timescales, and oceanic thermal re-emergence (i.e., the annual re-emergence of temperature anomalies temporarily obscured by the summer mixed layer). Mantua et al. (1997; see also An et al. 2007, Park et al. 2003, or Hartmann, 2015) recognized that the ENSO and PDO patterns are related, both spatially and temporally, leading to suggestions that the PDO may only represent the extratropical time integral of ENSO forcing (Zhang et al., 1997). However, some of the effects of these two oscillations are separable; for instance, there is no ENSO signal in Alaskan salmon harvests data (Mantua et al. 1997), and there are differences in Western USA precipitation depending on the combination of the (positive and negative) phases of the PDO and SOI indices (Wise, 2010). Goodrich (2007) uncovered PDO-related changes in winter precipitation and drought in years when the ENSO phase was neutral. Other distinctions are noted in Cha et al. (2018) and Wills et al. (2019).

The PDO continues to be an active area of research. Recent results of interest include those of Cha et al. (2018), who document relationships to the PDO of sea levels and trade winds in the equatorial Pacific, and those of Mills et al. (2019), who traced the influence of an intensified Aleutian Low (during the PDO warm (positive) phase) on the subpolar gyre and on the Kuroshio-Oyashio Extension in the North Pacific.

6.5. Coupled oscillations and the 18.6-yr lunar nodal cycle

Signatures of ENSO and of the PDO are robustly expressed in drought atlases for North America, for Monsoon Asia, and for the Old World (Baek et al., 2017). This strongly suggests that the coupled oscillations and drought must be physically connected in some way. However, as detailed above, it has not yet been possible to identify atmospheric or oceanic or coupled system factors that *control* drought occurrence. Specifically, for the Western USA, there are correlations of drought occurrence with cold SSTs in the eastern Pacific, and also with the negative phases of the ENSO and PDO oscillations, but some droughts (such as that of 2011-2015) have occurred in the absence of any apparent SST forcing (Seager et al., 2015). Thus, our existing conceptual and physical models for drought occurrence still appear to be deficient in some way.

Yasuda (2009) employed multiple statistical techniques on two samples of reconstructed PDO data (based on tree-ring chronologies) to uncover a statistically significant 18.6-yr lunar nodal cycle component within the Pacific Decadal Oscillation. In subsequent numerical simulations, Tanaka et al. (2012), Osafune et al. (2014), and Osafune et al. (2020) confirmed that the 18.6-yr cycle "determines the spatial structure of the PDO."

A robust relationship linking the December-February mature-phase ENSO cycle and the 18.6-yr lunar nodal cycle is described in Yasuda (2018). El Niño events and La Niñas occur preferentially at particular phases of the 18.6-yr cycle. In an earlier study, Cerveny & Shaffer (2001) found that equatorial Pacific SSTs and the Southern Oscillation Index (SOI) were each significantly correlated with maxima of the lunar declination (Fig. 7).

The above findings take on added significance in juxtaposition with the orbit-spin coupling mechanism described here. Previously, given the well-known quantitative deficiencies of tidal mechanisms, the many correlations found linking atmospheric and oceanic indices with the 18.6-yr lunar nodal cycle (Sections 4.2.1 and 6.2) could be dismissed as curiosities. That is no longer the case, as there now exists a cogent physical mechanism capable of exciting a quantitatively significant Earth system response at the 18.6-yr LNC period (Sections 2, 4 and 5 above).

Atmospheric modes of oscillation and coupled ocean-atmosphere modes of oscillation represent the integrated sum of all relevant effects and forces acting on the system studied. One of the goals of principal components analysis is to gain insight on the underlying physical processes. However, it is also widely acknowledged that "EOF analysis is a statistical representation of the variability of the climate system that might not reliably represent the underlying dynamics" (cf. Lee et al., 2019).

We submit that the documented relationships of the 18.6-yr LNC to ENSO (Yasuda, 2018) and the PDO (Yasuda, 2009) may now be interpreted as evidence of a relationship between 1) a driving *physical* process (i.e., the orbit-spin coupling torque associated with the LNC; § 4.2), and 2) a linked pair of *statistically defined* modes of coupled Earth-system variability. We conjecture that the EMS component of the orbit-spin coupling torque on the Earth system may contribute to the excitation of both ENSO and the PDO.

We are led to ask: how will modern AOGCMs respond to forcing by orbit-spin coupling torques on ENSO and PDO timescales? We conjecture that coupled system oscillations similar to ENSO and the PDO may emerge naturally within the simulations of the driven models.

To complete our discussion of ocean-atmosphere coupling, we reiterate that the contributions to bidecadal Earth system variability of the EMB component (Section 4.1) have been neglected here. The EMS component, with its stable 18.6-yr phase and amplitude modulation (Fig. 5b), can only represent one part of the full story, in connection with the occurrence of Western USA droughts and the excitation of Earth system variability. Instead, we must look to the resultant orbit-spin coupling torque of Fig. 8 to drive the system. Even then, we are likely to find that antecedent conditions within the system play an important role in the determination of forecasted outcomes in our simulations.

## 7. Forecasting future states of the atmospheric circulation: The late 2020s

7.1. Solar system dynamics

In this Section, we look more closely at the future evolution of the dynamical (solar) system and the associated variability of the orbit-spin coupling torque for the years 2022-2030. A polar plot of the solar motion for the years 2013-2052 is provided in Fig. 9, continuing the sequence presented earlier in Fig. 6. The star symbol identifies the Sun's position, along this trajectory, at the start of 2022. The current large displacement of the Sun's center from the solar system barycenter is consistent with the large present amplitude of the torque, as illustrated in Fig. 8. The Sun's orbital velocity along this trajectory will slow considerably as it approaches peribac in 2030.

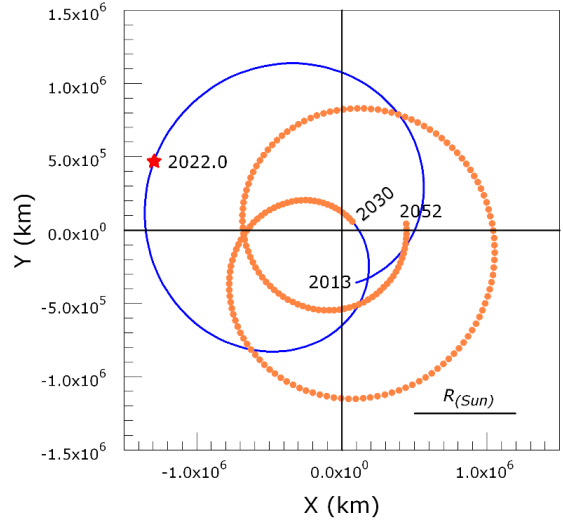

**Figure 9**. Counterclockwise (prograde) trajectory of the Sun's center with respect to the solar system barycenter for the years 2013-2052. A red star indicates the Sun's position on 1 January 2022.

The magnitude of the orbit-spin coupling torque on the Earth system in the years 2020-2030 is illustrated in Fig. 10, which is a zoomed-in version of a portion of the corresponding time series of Fig. 8. Figure 10 may additionally be compared with Fig. 3, which includes the year 2020 only. Peak magnitudes of the torque become progressively smaller during the time span illustrated, while the prominent quasi-semiannual modulation (previously discussed in connection with Fig. 3) becomes less and less well-defined.

Statistics of the torque amplitude for the peak amplitude years 2020-2023 were compared with the corresponding (mean and standard deviation) torque amplitude values for the California drought years 2011-2014 in Section 5 above. Substantial differences were found in that comparison. Here we compare the values for the 2020-2023 high torque interval (mean = 163.6, σ = 106.6) with those for 2029-2032 (mean = 97.4, σ = 51.7). As in the earlier comparison, we find substantial reductions in both metrics, for the 2030 torque minimum, as compared with the high-torque-amplitude interval 2020-2023. Next, comparing values for the bracketing torque minima, we find similar mean amplitudes for the two periods (95.4 for 2011-2014, vs 97.4 for 2029-2032). The torque amplitude for the latter period exhibits substantially less time variability, however, as measured by the standard deviation; σ for 2011-2014 was 62.0, while σ for 2029-2032 is 51.7.

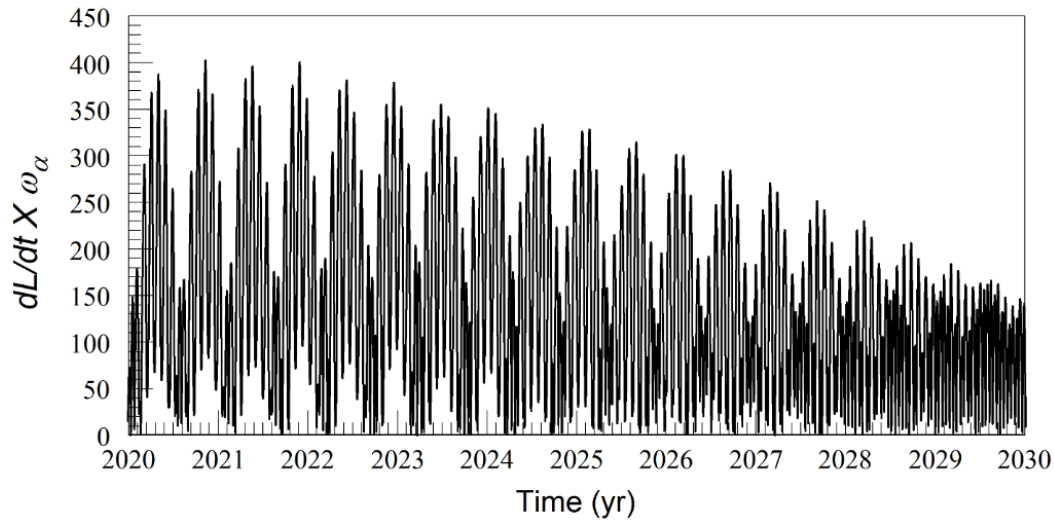

**Figure 10**. Magnitude of the Earth system torque for the years 2020-2030. Units are as in Figs. 5 and 8.

Figure 10 additionally serves to illustrate how the phasing of the torque waveform changes progressively with time with respect to the annual cycle of the seasons. This may be seen by noting the changing position of the local minima along the x axis in the years 2020-2025. The local minimum at the start of 2021 has shifted later in the year, in comparison with that seen at the start of 2020; similarly, those for early 2022 and 2023 are progressively shifted later in the year. The incommensurability of cycle periods, with respect to the annual cycle, together with the amplitude modulation, gives rise to inter-annual variability (as documented previously in the case of Mars (Mischna & Shirley, 2017).

An unusually high level of weather and climate variability was recorded in 2020. To cite just one example: More named Atlantic hurricanes were recorded in 2020 than in any other year of the historic record. We note in passing that peak torque amplitudes in 2020 were higher, albeit by a small margin, than in any other year of our study interval (Fig. 8). With reference to Fig. 10, it is apparent that we will continue to experience high peak torque levels (≥ ~350 in the units of Figs. 5, 8, and 10) for the next several years. Peak torque levels will be reduced thereafter, approximately by a factor of 2, between 2025 and 2030. As will be discussed later in Section 8, the current high levels of torque may represent a complicating factor for attribution studies seeking to uncover relationships between increasing greenhouse gas forcing and short-term weather and climate extremes (IPCC 2021, Chapter 3).

An expanded view of Earth system torques for the years 2028-2030 is provided in Fig. 11. These annual plots may usefully be compared with that for 2020 provided in Fig. 3a. We first note the strong similarity between the records for 2028, 2029, and 2030. Amplitudes exhibit relatively little time variability in these years, especially in 2030. The phasing is much more regular than in 2020; each year includes ~25 pulses, which correspond to the fortnightly cycle discussed earlier in Sections 3.1.4 and 3.1.5. As indicated in Fig. 3a, during high peak torque intervals, pulse durations are much less regular.

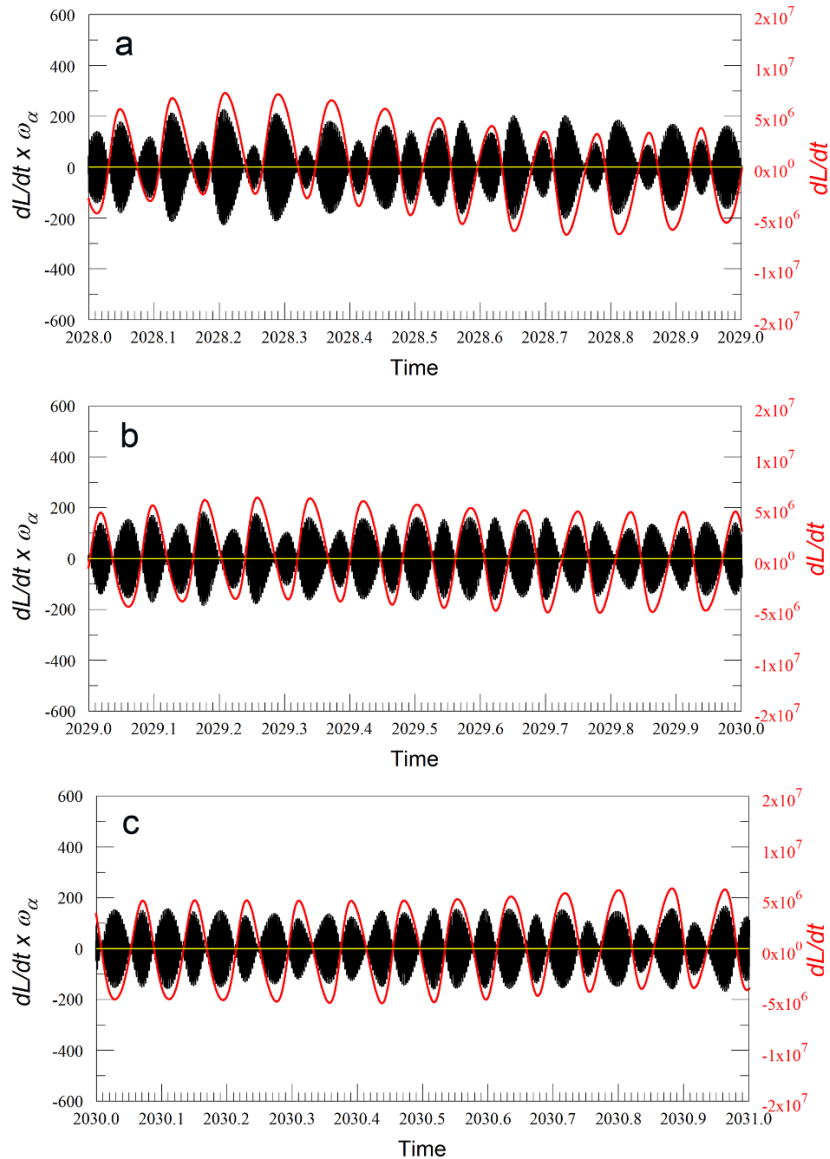

**Figure 11**. Orbit-spin coupling torques on the Earth system in the years 2028-2030. Units are as in Figs. 3, 5, 8, and 10 above. a): 2028. b): 2029. c): 2030. The fortnightly cycle arising from the EMS component dominates both in the forcing function *dL/dt* and in the torque *dL/dt* × *ω$_α$*.

We conclude from Fig. 11 that the variability of the torque in these years is almost entirely due to processes occurring within the Earth-Moon system. The Earth-Moon barycenter (EMB) component nearly disappears in these years, as the Sun's motion (Fig. 9) slows and its angular momentum about the solar system barycenter is minimized. The dominance of the fortnightly cycle in the Earth system torque in the years 2028-2030 raises questions of the nature of the atmospheric response during these years. Many past investigations have uncovered fortnightly cycles in precipitation and other atmospheric indices (Section 3.1.5). It will be of interest to see if signatures of the fortnightly cycle may emerge more or less strongly in weather data during the years illustrated in Fig. 11.

7.2. A prediction: Reduced time variability of the large-scale circulation in the late 2020s

The solar radiative input to the terrestrial atmosphere will exhibit the usual <3% annual modulation in coming years, but the solar energy input to the system should not otherwise vary significantly over the time period illustrated in Figs. 8-10. The upcoming solar peribac of 2030 thus provides an important opportunity for calibrating the response of the terrestrial atmosphere to variable forcing by orbit-spin coupling.

The upcoming reduction in the magnitude of the torque, by a factor of ~40%, relative to current values, as documented for the end of this decade in the present study, should be accompanied by an observable reduction in the variability with time of the large-scale circulation. Global solutions for parameters such as the total energy and kinetic energy of atmospheric motions on timescales of days to months should capture and quantify diminished (less energetic and less variable) weather and climate effects of pulsed additions of momentum to the circulation of lesser magnitude. We consider this to represent a critical experiment from a pure physics perspective. If the time variability of the large-scale circulation of the terrestrial atmosphere during 2029-2031 is indistinguishable from that of 2020-2023, this will argue against a significant role for orbit-spin coupling in the generation of terrestrial weather and climate variability.

7.3. Forecast: An extended Western USA drought beginning in 2028 ± 4 yr

Every close approach by the Sun to the solar system barycenter since the US Civil War and the administration of President Abraham Lincoln has been accompanied by a multiyear drought episode in the Western USA (Table 2 and Fig. 8). We have no reason to believe that the solar peribac of 2030.0 will differ from the prior eight such episodes in this respect. The upcoming torque minimum (Fig. 8), centered on the peribac passage of 2030, is the deepest such minimum falling within our study period.

We expect that anomalous stationary Rossby wave patterns and strong Eastern Pacific ridging (as described in § 5.1 and § 5.2 above) will occur more frequently and become more persistent in some or all of the years immediately previous to and following the solar peribac of 2030. However, our present state of knowledge is insufficient to allow more definitive statements to be made at this time. In particular, we cannot yet be sure that accompanying ENSO and/or PDO negative phase SST conditions will closely mirror the patterns derived from analyses of past events. We additionally lack information sufficient to estimate or assess the intensity of the forecasted event.

We can constrain the timing of the forecasted multiyear Western USA drought episode by means of a small-numbers statistical analysis of the data already presented above in Table 2. We first obtain the differences in time between the years of drought inception and the times of solar peribac. We obtain the following time differences, in years: [3.6, 1.2, -0.7, 0.7, 5.4, 0.2, 3.3, 2.8]. Here a positive value indicates that the inception date precedes the peribac date. The signed mean difference for this series is 2.1 yr, with standard deviation (σ) = 2.05 yr. The forecast (centroid) multiyear drought episode inception date then becomes (2030.0 – 2.1) = 2027.9, with ±2-σ bracketing times yielding a range of dates from 2023.8 to 2032.0. Given the 1-yr resolution of the drought episodes initiation dates of Table 2, we round the above estimates (eliminating the decimals), obtaining a drought forecast inception date of 2028, with standard deviation σ=2 yr.

Given the significant economic, ecological, and social impacts of multiyear droughts of the Western USA, two implications come immediately to mind:

1). Future improvements in the precision of this relatively crude forecast may yield considerable added value for water resource managers and other stakeholders.

2). A "wait and see" approach to forecast evaluation clearly cannot be justified as an optimal near-term strategy for moving forward with follow-up investigations.

The above considerations underscore the pressing need for an immediate effort to falsify or confirm the applicability of the orbit-spin coupling hypothesis in the terrestrial case. Targeted numerical simulations with GCMs, as already accomplished for Mars, arguably represent the most expeditious approach to validation. If the efficacy of the coupling is confirmed, subsequent simulations should enable substantial improvements in the temporal precision and spatiotemporal resolution of forecasts. In this connection, we reiterate that orbit-spin coupling based sub-seasonal timescale forecasts of future episodes of large-scale atmospheric instability on Mars are already available (Shirley, McKim, et al., 2020). Development of a similar or better capability for terrestrial forecast purposes is a worthwhile goal.

7.4. Decadal climate prediction: The present state of the art

Decadal climate prediction addresses the vulnerabilities of societies in terms of food security, water resource availability, energy supply, and populations migration and conflict (Meehl et al., 2014; Cassou et al., 2018; Smith et al., 2019; IPCC, 2021). Important advances in decadal prediction skill have been made in the past few years (Smith et al., 2019). Thus, it is germane to briefly review present methodologies of decadal timescale climate prediction in juxtaposition with the forecast obtained in the present investigation.

Decadal prediction skill is much enhanced through model initialization (Meehl et al., 2014; Smith et al., 2019), typically by including observed SSTs, which help determine heat uptake and redistribution within the ocean-atmosphere system. The initial state may contribute to the skill of predictions over the first few years. In general, the current models produce better predictions of temperatures than of precipitation, or of the state of the large-scale circulation (Smith et al., 2019).

Atmospheric oscillations (such as ENSO, and the PDO, as discussed above) can be considered to comprise a system of teleconnections (Cassou et al., 2018; Steptoe et al., 2018). The persistence of the oscillations, and their teleconnections, on climate-related timescales, can contribute to improved decadal timescale climate prediction. However, the temporal non-stationarity of the oscillations introduces corresponding uncertainties in the predictions. Probabilistic statistical methodologies become necessary when one thinks in terms of "broad spectrum variability instead of temporal oscillations with preferred timescales" (Cassou et al., 2018).

Chaotic dynamics and internal variability, as well as greenhouse gas warming effects (to be discussed in Section 8 below), further impact decadal timescale predictions. These factors may be addressed through the statistical analysis of large ensembles of climate model runs (Smith et al., 2019). Changes in the frequency of extreme events are thought to be predictable through methodologies such as these.

To properly contrast the current decadal timescale prediction capability with the forecast made here, it may be helpful to recall some of the acknowledged deficiencies of the decadal prediction methods. Smith et al. (2019) note the following three deficiencies: 1) The numerical models typically underestimate the observed amplitude of variability; 2) "the relationship between surface temperatures and atmospheric circulation anomalies is not simulated correctly by the initialized forecasts"; and 3) "the modelled internal variability is not aligned, on average, with that of the real world." From these and other considerations, Smith et al. (2019) conclude that "there is a clear deficiency in climate models that urgently needs to be addressed."

The hypothesis of a deterministic forcing of the large-scale circulation by torques due to orbit spin coupling may potentially explain and resolve each of the above deficiencies. Momentum added to the circulation by orbit-spin coupling torques must be expected to increase the amplitude of observed variability (deficiency 1), in comparison with unforced simulations. The added torques must likewise be expected to significantly perturb nominal relationships between surface temperatures and circulation anomalies (deficiency 2).

The third of the cited deficiencies (the temporal misalignment of modelled internal variability with that of the real world) is fundamental. Sections 3-5 demonstrate that forcing by orbit-spin coupling is strongly deterministic, while exhibiting marked temporal variability (Fig. 8). If orbit-spin coupling is operative within the terrestrial system, then we would expect that the outcomes of coupled system models *lacking* this forcing should bear little relationship to the Earth system variability observed in the real world. In this connection, it is worthwhile to recall that a close correspondence between forced model outcomes and real world planetary-scale weather and climate anomalies has already been demonstrated for Mars (Mischna & Shirley, 2017; Shirley, Newman, et al., 2019; Newman et al., 2019; Shirley, Kleinbδhl et al., 2019; Shirley, McKim, et al., 2020).

The decadal climate prediction strategies and methods described in Meehl et al. (2014), Cassou et al. (2018), and Smith et al. (2019) are all probabilistic in nature. While we have employed statistics here, in order to estimate a likely range of inception dates for our forecast, the reader should recognize that the underlying approach is *not* based on statistical probabilities. Our prediction and forecast is instead based on deterministic physical processes and interactions. The lead time for the forecast presented here is approximately 6 years. Purely probabilistic approaches to decadal climate prediction are unlikely to yield forecasts of significant future climate anomalies with comparable lead times, at least in the near future. Optimal performance in decadal climate prediction may conceivably require the addition of solar system dynamical forcing information to existing decadal prediction models and systems.

## 8. Discussion

Known causes of ongoing and future terrestrial climate change may be usefully separated into two categories. These are: 1) anthropogenic influences on future climates, and 2) natural variability. According to the IPCC (2021), the latter may be defined as "climate fluctuations that occur without any human influence, that is, internal variability combined with the response to external natural factors such as volcanic eruptions, changes in solar activity, and on longer timescales, orbital effects and plate tectonics."

The nature of the cross-coupling between atmospheric natural variability and greenhouse gas warming is a topic of considerable interest and current research activity. In what ways may the present results bear upon these questions?

8.1. Natural variability and chaotic dynamics

8.1.1. Natural variability

Natural variability is a catch-all phrase describing a wide spectrum of Earth system phenomena whose underlying causes are either unknown or are plausibly chaotic. Interannual, seasonal, and shorter timescale atmospheric variability (for instance, of precipitation, or temperature) is often characterized as resulting from natural variability. Large-scale coupled oscillations of the Earth system, such as ENSO and the PDO, may also be categorized as forms of natural variability.

The 77 peer-reviewed investigations documenting the presence of lunar cycles in observations of atmospheric and oceanic indices cited above in Sections 3.1.5, 4.2.1, and 6.5 likewise comprise investigations of Earth system natural variability. Multiyear droughts of the Western USA constitute a form of natural variability of weather and climate. The present investigation may thus accurately be characterized as an exploration of the physical origins of natural variability within the coupled ocean-atmosphere system.

8.1.2. Chaotic dynamics

Chaotic dynamics (Lorentz, 1963, 1969, 1984) presently provides our consensus best explanation for atmospheric natural variability, i.e., for the observed complex time variability of the terrestrial atmospheric circulation on timescales from days (Bauer et al., 2015; Mariotti et al., 2018) to decades (Du et al., 2012; Smith et al., 2019). The governing equations describing the evolution of atmospheric weather and climate (Holton, 1992) exhibit a sensitive dependence on initial conditions (a hallmark of chaotic dynamics). Sensitive dependence on initial conditions in the system of equations employed by Lorentz (1963) leads to the conclusion that details of future weather cannot accurately be predicted, even in principle, much beyond 2 weeks in the future (Lorentz, 1969; Bauer et al., 2015).

The applicability of chaotic dynamics in connection with studies and behaviors of atmospheres rests upon one crucial underlying assumption. This is *the assumption that nothing of any importance has been left out of the system of equations used to describe the physical system*. We are led to ask: Would current AGCMs and AOGCMs, when supplementally driven by the torque of equation (3), exhibit a similar sensitive dependence on initial conditions? That is, would the calculated outcomes of ensembles of *cta*-driven models (each initialized with slightly different starting conditions) rapidly drift apart with the passage of time? Or would these exhibit a greater measure of consistency between realizations than is seen with current simulations, as a consequence of the highly deterministic forcing of the external torque given by equation (3)? The answer to this question may best be obtained by running new ensembles of forced and control simulations, with and without dynamical forcing by orbit-spin coupling, in each case with suitably modified initial conditions. High priority should be given to new experiments addressing this question.

Modern numerical modeling investigations addressing the forward evolution of the atmospheric system typically employ an ensembles approach to characterize the range of possible outcomes arising due to a sensitive dependence on initial conditions. Many independent model runs, or realizations, each starting from slightly different initial conditions, are analyzed statistically to obtain an appropriate mean system forward projection or forecast. This approach has led to significant improvements in forecasts (Bauer et al., 2015; Meehl et al., 2016; Kirtman et al., 2017; Scaife & Smith, 2018; Mariotti et al., 2018; Zhang & Kirtman, 2019; Smith et al., 2019; Alley et al., 2019).

We nonetheless recognize an inherent problem with this approach. The issue derives from the above-noted fundamental underlying assumption of the chaotic dynamics paradigm for atmospheric variability. We must seriously consider the possibility that all such simulations may have been obtained using *an incomplete physical specification of the system*. If this is the case, then even a very large number of realizations may fail to fully characterize the true range of internal (natural) variability of the complex system under investigation. In this connection, it may be helpful to recall the documented range of natural variability of the pre-industrial climate during the past millennium, which includes both the Medieval warm period (~900-1250 CE) and the Little Ice Age (~1550-1850 CE) (Lamb, 1977). The former period includes several Western USA drought episodes of more than one decade in duration (i.e., megadroughts) which severely impacted the agricultural societies of the time (Woodhouse & Overpeck, 1998; Fye et al., 2003; Cook et al., 2004, 2007, 2010; Woodhouse et al., 2010; Appendix 2).

8.2. Greenhouse gas warming and natural variability

The 2021 IPCC Assessment Report 6 (Climate change 2021: The Physical Science Basis) is a monumental collaborative achievement, with contributions spanning many science and technical disciplines and communities. The Report (AR6) comprehensively describes the known weather and climate impacts of increasing atmospheric greenhouse gases (GHG). AR6 demonstrates the reality of an ongoing anthropogenic influence on important global climate metrics, including global mean temperature, ocean surface temperatures, polar sea ice cover, and other indices. AR6 will usefully inform decision makers and stakeholders for years to come. It is beyond the scope of this paper to comprehensively discuss any portion of AR6. It is appropriate, however, to draw the reader's attention to certain subject areas wherein the mechanism considered here may contribute to an improved understanding of the global warming problem.

One of the key findings highlighted in the Summary for Policy Makers (IPCC, 2021) is that "natural drivers and internal variability will modulate human-caused changes, especially on regional scales and in the near term." The projected effects of GHG forcing may be "either amplified or attenuated by internal variability" (at high confidence). IPCC (2021) concludes, with many prior teams and investigators (Woodhouse & Overpeck, 1998; Dai et al., 2013; Trenberth et al., 2013; Cook et al., 2004, 2010, 2015; Woodhouse et al., 2010; Cook et al., 2019), that global warming may be expected to amplify drought severity in the future. A number of investigations report detection of an influence of GHG warming on drought occurrence (Wang et al., 2014; Seager et al., 2014; Williams et al., 2015); however, in most of these cases, a proportionally larger role for natural variability is also noted.

Atmospheric natural variability introduces uncertainty in projections, forecasts, and predictions of future weather and climate states (IPCC 2021, § 1.4.3.1). In the conventional view, from a high-level perspective, natural variability represents a sort of "black box," in which known inputs are nonlinearly scrambled to yield unpredictable outcomes. The uncertainty due to internal variability increases for small spatial scales. Investigators thus early on recognized that the most useful and reliable metrics for characterizing GHG effects must be global in scale (such as the global mean surface temperature, GMST) (cf. Schlesinger, 1994). Global integration reduces the influence of internal natural variability, and thus the total uncertainty becomes dominated by emissions scenario and model response uncertainties (IPCC 2021, § 1.4.3.2).

Under the chaotic dynamics paradigm, as outlined above (§ 8.1.2), investigators of GHG warming effects are free to equate the true range of natural variability with the range of outcomes of ensembles of simulations of varied starting conditions. In general, however, simulations performed to study GHG effects cannot be employed to identify *trends* of *natural* variability. In light of the known broad spectrum of weather and climate variability (Lamb 1972, 1977), this clearly represents a shortcoming of existing climate modeling methodologies. Trends of natural variability can impact GMST records. In particular, the cold phase of the Interdecadal Pacific Oscillation appears to have played a role in the genesis of the hiatus of global warming that was observed from ~1998-2013 (Meehl et al., 2016). (Earlier studies highlighting interactions between atmospheric and ocean dynamics and GHG abundance or forcing include those of Bacastow (1976), Newell & Weare (1977), and Bacastow et al. (1980).

The IPCC RP6 (2021, § 1.4.3.2) offers hope that advances in decadal prediction may lead to a narrowing of the uncertainties in the trajectory of the climate for a few years ahead. However, at present, the temporal phasing of observed modes of natural variability such as ENSO and the PDO cannot reliably be predicted.

The time history of the Earth system torque of Fig. 8 likely constitutes a forcing mechanism for some portion of the observed natural variability of the atmosphere that is presently attributed to chaotic dynamics. If so, then it may be possible, in future targeted investigations, to disentangle dynamically forced natural variability and superimposed GHG warming effects. An improved calibration of GHG warming effects may emerge from such an exercise. This may conceivably lead to tighter constraints on the sensitivity parameter, and similarly to reduced uncertainty in forecasted future temperature trends, i.e., in the magnitude and timing of future global warming. Prospective improvements in these areas could by themselves be of sufficient economic value to justify a program of comprehensive investigations of orbit-spin coupling effects on the Earth system.

8.3. Next steps

Most fundamentally, we wish to determine whether the level of agreement between terrestrial weather and climate model outcomes and observations may be improved through the use of a nonzero value of the coupling efficiency parameter $c$ of equation (3) in numerical simulations. Prior investigations of the effects of orbit-spin coupling on the Martian atmosphere confirm that the use of a nonzero $c$ value yields non-trivial improvements in the temporal agreement between model outcomes and observations for that planet (Shirley & Mischna, 2017; Mischna & Shirley, 2017; Newman et al., 2019; Shirley, Newman et al., 2019; Shirley, Kleinböhl, et al., 2019; Shirley, McKim, et al., 2020). Numerical investigations that address the question of zero-versus-nonzero $c$ values for reproducing terrestrial weather and climate variability and trends should thus take precedence in any phased program of testing. Mischna & Shirley (2017) outline a methodology based on comparisons of forced and control model runs.

Incorporation of the orbit-spin coupling accelerations (*cta*) within existing GCMs may not require major investments in coding or revisions to program structures. Mischna & Shirley (2017) modified the dynamical core of the Mars version (Richardson et al., 2007; Toigo et al, 2012) of the Weather Research and Forecasting (WRF) GCM (Scamarock & Klemp, 2008) to include added incremental wind velocity components in the *u* and *v* directions (due to the coupling). The velocity increments were obtained, for each grid point and time step, by integrating the calculated dynamical accelerations for times corresponding to each internal time step of the model. The accelerations were obtained from data files similar to those employed to produce Figs. 3, 4, 5, 10, and 11 of this paper. The 2-hr time step accelerations data employed for the present investigation are archived at (TBD) and are freely available. Algorithms for calculating the accelerations for other time intervals and with other time steps are likewise published and freely available.

On the basis of the results described in Section 5, on the basis of the scale analysis of Section 2.3, and on the basis of the dozens of prior investigations reporting the presence of unexplained dynamical signals in atmospheric indices (Sections 3.1.5 and 4.2.1), we expect that the effects of the torque on the terrestrial atmosphere are unlikely to lie at or near the limits of detection (when comparing forced model outcomes with control runs).

### 8.3.1. Weather timescale investigations

We recognize the existence of a wide range of viable test procedures and approaches to hypothesis testing. A new working hypothesis for explaining the variability of the atmospheric circulation on weather-related timescales was introduced in Section 3. Fortnightly timescale variability of the torque is illustrated in Figs. 3, 4, and 11. Many prior studies find relationships linking precipitation with fortnightly cycles (Section 3.1.5). Numerical weather forecasting models incorporating the *cta* could thus directly address questions of possible relationships linking precipitation and heavy rainfalls with the fortnightly modulation of the torque. Weather timescale investigations could also revisit the question of a theoretical limit on predictability due to sensitive dependence on initial conditions. Weather timescale investigations may additionally address the question of the proportion of observed atmospheric natural variability (at these timescales) that may be attributable to forcing by orbit-spin coupling torques.

### 8.3.2. Interannual and longer timescale investigations

Experiments addressing relationships of orbit-spin coupling torques and interannual to bidecadal atmospheric variability, as in connection with multiyear droughts, are also needed. Somewhat greater resource investments may be required to perform hypothesis testing employing coupled system models over timescales of decades to centuries. Adaptation of existing coupled models to include the orbit-spin coupling torques may yield significant benefits. For instance, in a recent discussion of Coupled Model Intercomparison Project-5 (CMIP5) model performance, Lee et al. (2019) make two cogent observations. In connection with the representation by models of atmospheric modes such as the North Atlantic Oscillation, the PDO, the Southern Annular mode, and others, these authors first encouragingly note that "in many cases, models are doing a credible job at capturing the observationally-based estimates of patterns". Lee et al. (2019) subsequently observe, however, that "*there is no reason to expect the simulated internal variability to be in-phase with the observational record.*" The inclusion of orbit-spin coupling torques in future simulations may substantially improve the agreement of model outcomes with terrestrial observations, in the time domain, as already demonstrated for Mars (Mischna & Shirley, 2017; Newman et al., 2019; Shirley, Newman et al., 2019).

Separate determination of optimal values of the coupling coefficient *c* for the oceanic and atmospheric components of coupled models (Appendix 1) will likely be required for such investigations. Just as the properties of the media influence the fall rates of freely falling bodies in air and water, the response of the oceans to the *cta* is likely to differ from that of the atmosphere, under conditions in which both experience an identical forcing acceleration. Coupled (ocean-atmosphere) models with separately calibrated *c* values may offer an optimal path forward for evaluating the working hypothesis for multiyear Western USA drought occurrence introduced in Section 5, and thereby for achieving an iterative refinement of the precision of the forecast given here in Section 7.3.

8.3.3. Initial testing and GHG

While the combined effects of GHG forcing and natural variability in the form of solar system dynamical forcing may be of primary importance in socioeconomic terms, initial proof of concept investigations need not explicitly include or address effects due to long term secular changes in GHG forcing. Investigations holding constant the abundance of atmospheric GHG may more precisely isolate the effects of the orbit-spin coupling torques on the large-scale circulation, and may more precisely constrain the $c$ value (Appendix 1) that is most appropriate for the timescale and phenomena being investigated.

Many other approaches to hypothesis testing (in addition to those listed in this Section, and those described previously in Sections 3.1.5, 5.2, 7.2, and 6.5) are possible and should be encouraged.

8.3.4. Cost-benefit relationships

"Predictability at decadal and longer time scales would considerably improve decision making and risk management in agriculture and therefore make an important contribution toward sustainable development… At regional scales such forecasting ability would also assist with the implementation of flood and drought mitigation strategies and programs." (Meinke et al., 2005).

The cost to comprehensively test the orbit-spin coupling hypothesis pales in comparison with both 1) annual US government crop insurance payouts, and 2) the average annual mitigation costs of weather-related natural disasters. Each of these is measured in the billions of US dollars per year. The next multiyear drought of the western USA is likely to engender costs of this same magnitude. For this reason, we strongly advocate rapidly establishing funding programs to support numerical modeling investigations specifically designed to explore and constrain the role of orbit-spin coupling in the generation of atmospheric natural variability.

## 9. Summary and Conclusions

We analyze the time history of orbit-spin coupling torques on the Earth system from 1860-2040 to assess the applicability of the orbit-spin coupling mechanism for investigations of terrestrial weather and climate variability. Under the orbit-spin coupling hypothesis, planetary atmospheres participate in a transfer of angular momentum between the reservoirs of the planetary orbital and rotational motions. The coupling takes the form of a reversing torque, with the axis of the torque lying within the planetary equatorial plane. Pulses of momentum are episodically added to the circulation of the atmosphere by this mechanism. To obtain a current best estimate of the torque peak magnitude, we employ a coupling efficiency coefficient that was previously optimized in studies of the Mars atmosphere (which lies within the same regime as the Earth). We obtain (local) peak coupling term accelerations (*cta*) of up to ~1.2 mm s$^{-2}$, which are large enough to give rise to important effects within the large-scale circulation of Earth's atmosphere. The torque is highly deterministic and is easily calculated, allowing for unambiguous hypothesis testing.

Ocean-atmosphere interactions, coupled ocean-atmosphere modes of variability, and relationships between atmospheric natural variability and greenhouse gas forcing are each briefly discussed here in juxtaposition with orbit-spin coupling effects. While we primarily focus on decadal to bidecadal timescale processes, our investigation has also uncovered possible modes of deterministic external forcing operating on weather-related timescales.

9.1. Principal findings

Results and principal findings of this investigation pertaining to atmospheric variability on decadal and longer timescales include the following:

1. Our investigation has uncovered a one-to-one correspondence linking orbit-spin coupling torque minima with the occurrence of multiyear droughts of the Western USA. Every close approach by the Sun to the solar system barycenter since the US Civil War has been accompanied by a multiyear drought episode in the Western USA.

2. Orbit-spin coupling for the first time provides a quantitatively viable physical mechanism to explain the well-documented 22-yr rhythm of drought occurrence in the Western USA.

On the basis of the above findings, we present a new working hypothesis for the occurrence of multiyear episodes of drought in the Western USA associated with times of close approach of the Sun to the solar system barycenter. We propose that the weakly varying large-scale circulation patterns associated with prolonged droughts arise at times when the orbit-spin coupling torque magnitude and its time variability are strongly reduced and are approaching minimum values.

Drought occurrence in the Western USA also exhibits a statistically significant 18.6-yr recurrence tendency that has previously gone without an acceptable physical explanation. We submit that:

3. Orbit-spin coupling provides, for the first time, a quantitatively viable working hypothesis for the excitation of spatially widespread atmospheric bidecadal variability at the 18.6-yr periodicity of the retrograde revolution of the lunar nodes.

Results and principal findings of this investigation pertaining to atmospheric variability on monthly, fortnightly, and shorter timescales include the following:

4. Orbit-spin coupling torques exhibit a pronounced amplitude modulation that is associated with the fortnightly and monthly orbital cycles of the Earth and Moon within the Earth-Moon system.

We introduce a working hypothesis for the excitation of atmospheric variability on timescales from days to weeks, wherein the torque imparts pulses of momentum that perturb the large-scale circulation. We have shown that the torque exhibits substantial amplitude modulation on fortnightly and lunar monthly timescales. The torque in addition exhibits complex variability in phasing. We conjecture that the forcing mechanism proposed may account for the results of many prior published investigations showing fortnightly to monthly timescale modulations of monsoon rainfall, atmospheric temperatures, atmospheric pressures, and precipitation. We additionally conjecture that the variability thus generated likely contributes to our inability to forecast weather variability beyond about 10 days in the future. We thus further submit that:

5. Orbit-spin coupling provides a quantitatively viable explanation for the fortnightly timescale atmospheric variability observed in globally distributed locations and in multiple indices.

9.2. Prediction

The upcoming orbit-spin coupling torque minimum of 2030 is the deepest such minimum within the 180-yr interval examined in this investigation. Torque magnitudes, and the time variability of the torque (as measured by the standard deviation of the magnitude values), will each be reduced substantially, in future years bracketing this episode.

Theoretical considerations and the results of the present analysis allow us to formulate the following predictive statement, which may be regarded as a critical test from a pure physics perspective: *If the time variability of the large-scale circulation of the terrestrial atmosphere during the low torque interval 2029-2031 is indistinguishable from that of the high torque interval 2020-2023, this will argue against a significant role for orbit-spin coupling in the generation of terrestrial weather and climate variability*. We predict that anomalously stable and persistent Northern Hemisphere Rossby wave patterns will develop, and be observed, in the years 2029-2031, as in prior multiyear drought episodes of the Western USA.

9.3. Forecast

Forecasting of future multiyear Western USA drought episodes is enabled by the deterministic nature of the external orbit-spin coupling torque on the Earth system. A future multiyear drought episode is expected to begin in $2028 \pm 4$ yr (where the quoted range corresponds to $\pm 2$ standard deviations). The centroid forecast inception date (2028) and its associated standard deviation (2 yr) have been obtained from a statistical analysis of a small number (8) of prior (historic) examples. Future numerical modeling investigations are likely to substantially improve upon both the precision and the uncertainty of the present forecast.

The present forecast has a lead time of ~6 yr.

9.4. Recommendations

We strongly advocate the initiation of a numerical modeling effort to explore and constrain the effects of orbit-spin coupling on the Earth system. (Similar efforts have already enabled successful subseasonal-timescale forecasts of episodes of large-scale atmospheric instability on Mars). An investment in numerical modeling is likely to substantially improve the precision and reduce the uncertainties of the present forecast. The economic, ecological, and social costs of multiyear droughts of the Western USA are substantial, with economic losses typically measured in the billions of US dollars. Improving the spatial and temporal resolution of the present forecast may enable better informed advance planning, and the implementation of targeted drought mitigation strategies, by resource planners, by the agricultural industry, and by other stakeholders.

**Appendix 1: Mars-Earth comparisons and the coupling efficiency coefficient *c***

We address three linked topics in this Appendix. We first make note of key physical similarities and differences of the atmospheric systems of Mars and the Earth. Through these comparisons we hope to identify factors that may impact or constrain each system's response to forcing by the coupling term accelerations (*cta*). Our second topic concerns the fundamental nature of *c* and the selection of an appropriate initial value of the coupling efficiency coefficient for terrestrial applications. In the final subsection, holding constant the coupling efficiency coefficient *c*, we obtain orbit-spin coupling torques on the Earth that are about five times larger than those previously obtained for Mars.

A1.1. Physical similarities and differences of the atmospheres of Mars and the Earth

The large-scale circulations of the atmospheres of Mars and the Earth are morphologically similar, with each showing strong overturning circulations (Hadley cells) in low latitudes, and strong westerlies in mid to higher latitudes, which are stronger in winter than in summer (Leovy, 2001; Barnes et al., 2018). Both atmospheres exhibit traveling baroclinic eddies (storm systems) in middle and higher latitudes, along with polar vortices that expand and shrink with the seasons. Rossby wave numbering schemes are employed to characterize large-scale circulation patterns in both cases.

Terrestrial GCMs such as the Weather Research and Forecasting (WRF) model (Skamarock and Klemp, 2008) have been successfully adapted for use at Mars (Richardson et al, 2007; Toigo et al., 2010), as the underlying physics (expressed in the governing equations) is similar in both cases. Mars GCMs (MGCMs) are adapted to implicitly account for many of the key differences between the two atmospheres, such as the average magnitude and annual cycle of solar irradiance, the difference in surface gravity, and the differences in the composition of the atmosphere. (Earth versus Mars ratios for these and other parameters are summarized in Table 2.1 of (Zurek, 2018). Radiative transfer schemes, model grid sizes, and topographic interactions may likewise be implemented similarly in Martian and terrestrial GCMs. MGCMs capably represent other factors, such as the large seasonal condensation of $CO_2$ ice from the atmosphere, and the much shorter thermal time constant at Mars. Reference data for the two atmospheres may be found in many published sources and online. An authoritative review of current knowledge of weather and climate on Mars may be found in the recent volume edited by Haberle and others (Haberle et al, 2018).

The angular velocity of the planetary rotation ($\omega_a$) appears in equation (3); for purposes of the present discussion, it is relevant to note that $\omega_a$ values are quite similar for Mars and the Earth (7.088 rad s$^{-1}$ for Mars, versus 7.292 rad s$^{-1}$ for the Earth), differing only by about 3%.

The atmospheric systems of Mars and the Earth *differ* in important ways as well. We mention three key issues here. A first important difference is found in the differences of the atmospheric mass and density for the two planets, which is reflected in a two-orders-of-magnitude difference in mean surface pressures. Whereas the mean atmospheric pressure on Earth is about 985 hPa (1 bar), on Mars the value is 6.9 HPa (.0069 bar). The density of the atmosphere is an important factor with reference to the coupling of the surface and atmosphere, through wind stress and frictional effects. Wind speed differences of the same magnitude on Mars and Earth produce substantially larger mountain torques and frictional losses on Earth, due to the larger mass and density of the accelerated atmosphere.

Atmospheric dust plays a much more important role at Mars than on Earth. Dust, once lifted into the atmosphere, reflects, absorbs, and re-radiates incoming solar energy, playing a role on Mars that is in some ways similar to that of water on Earth (Heavens et al., 2019).

Perhaps the most important difference between the atmospheric systems of Earth and Mars is found in the presence of abundant water, in the oceans and atmosphere, on Earth, as contrasted with the nearly complete absence of water on Mars. This leads to a number of differences that are of likely significance with respect to the response of the system to the *cta*. The terrestrial oceans have much higher heat capacity and thermal inertia than the dry rocky surface of Mars. As we have noted, the thermal memory of the oceans is important for explaining interannual variability of climates within the Earth system. The absence of oceans helps explain the large diurnal atmospheric temperature ranges observed on Mars.

The mechanical coupling between the atmosphere, the oceans, and the underlying solid Earth has long been of considerable research interest (cf. Oort, 1989). The analogous coupling at Mars, involving just the atmosphere and solid surface, is considerably less complex. The oceans may temporarily store and subsequently transmit momentum derived from the atmospheric motions, potentially adding a form of mechanical storage to the system, and possibly introducing a temporal lag affecting response times within the system.

A1.2. Nature, role, and estimation of the coupling efficiency coefficient *c*

The coefficient *c* characterizes the fractional portion of the orbital angular momentum that may participate in the excitation of geophysical variability. In the initial MGCM investigation by Mischna & Shirley (2017), a *c* value of $1.0 \times 10^{-12}$ was found to lead to unrealistically high Martian wind velocities; while a value of $1.0 \times 10^{-13}$ was insufficient to produce any notable differences from unforced (control) simulations. Thus a value of $5.0 \times 10^{-13}$ was adopted for that study, as described in detail in § 4 of Mischna & Shirley (2017).

Employment of a nonzero *c* value in numerical simulations can only be justified when this brings about a nontrivial improvement in the level of agreement between modeling outcomes and observations. While the use of an adjustable gain parameter like *c* may be faulted for being arbitrary, in some applications, experience has shown that in the present case, it is the *phasing* of the forcing function *dL/dt* that is most critical. Mischna & Shirley (2017) noted that no value of *c* could be expected to improve the agreement between modeling and observations, in the time domain, if the phasing of the forcing function with respect to the annual cycle was unfavorable.

Appendix E of Shirley (2017a) summarizes geophysical precedents for the introduction and use of *c*. There, *c* is first compared with the classical mechanical coefficient of friction introduced by Leonardo da Vinci. As in that case, the determination of the parameter value from first principles is a difficult task. In the present case, the underlying physical processes responsible for the coupling phenomena are yet to be fully understood (Shirley, 2017a). Thus, the empirical determination of $c$ in a variety of situations and circumstances should be given a high priority as a research objective. By experimentally constraining the value of $c$ in various situations, we may be able to more tightly constrain the nature and origins of the underlying causal mechanisms.

We recognize the possibility that $c$ may take on different values for different interacting subsystems of a subject body. The $c$ value determined in Mischna & Shirley (2017) does not represent an integrated global value, which could help characterize the level of associated rotational energy dissipation within the planet. Instead, its applicability is currently restricted to the problem of the momentum budget within the system of the atmosphere and solid surface. As indicated in the main text, we consider that the *cta* are applied directly to both the oceans and atmosphere; and we recognize the possibility that these systems may each individually respond somewhat differently (setting aside for a moment the phenomena arising at the interface between them). The other concentric shells of the planet similarly experience the acceleration field of Fig. 1, and these may likewise respond yet differently.

The $c$ parameter may additionally be found to depend on frequency, or on the mechanical or even the compositional properties of the interacting system components. In this way, $c$ is similar to the geophysical quality factor Q, which characterizes the amount of dissipation occurring during one cycle of an oscillation of a system. Different values of Q are routinely found for different subsystems of the Earth (additional discussion of this topic may be found in Appendix E of Shirley, 2017a).

The coefficient $c$ may thus be seen as a placeholder for quantifying the effects of an unknown number of weakly dissipative physical processes that macroscopically effect the coupling. Later investigations are needed to shed additional light on this problem.

From a practical standpoint, as noted in § 4.5 of Shirley (2017a), the upper bound for the $c$ parameter probably lies somewhere near $1.0 \times 10^{-12}$, as values much larger than this would be expected to degrade the precision of our solutions for the positions and velocities of solar system objects in ways that are not observed. It is possible that nontrivial geophysical effects could appear, in some situations, due to $c$ values below the threshold of $1.0 \times 10^{-13}$ found in Mishna & Shirley (2017). This is a question that needs to be explored. In any case, at the present time, when specifying a $c$ value of $5.0 \times 10^{-13}$ for the terrestrial case, it appears unlikely that we will be in error by much more than one order of magnitude.

Given the above uncertainties regarding the nature of the underlying physical coupling processes, together with the important differences of the surface-atmosphere coupling processes of the Martian and terrestrial atmospheric systems detailed above, we see no obvious way to formulate and justify the use of a modified $c$ value for the present investigation.

A1.3. Torque differences arising from solar system dynamics

The orbital angular momentum of the Earth (Table 1) is larger than that of Mars by a factor of ~7.5. In addition, the rates of change of the orbital angular momenta of Mars and the Earth are quite dissimilar, in both the character of their time variability, and in magnitude. To illustrate the difference in magnitude, in Fig. A1, we compare the time derivatives of the angular momentum (*dL/dt*) of the two bodies.

We note immediately in Fig. A1 both the larger amplitude of the terrestrial waveform and its higher frequency, quasi-monthly modulation, due to the presence of Earth's Moon. Due to the orbital motion of the Earth and Moon about their common barycenter (Fig. 2 and § 2.4), Earth's orbital angular momentum (with respect to the solar system inertial frame) is alternately greater than, and less than, its long-term mean value. Time variability with much lower frequency characterizes the waveform for Mars in Fig. A1. The peak-to-peak cycle times for both planets are slightly longer than one solar year, as defined for each subject body, due to the ongoing prograde orbital motion of the Sun with respect to the solar system barycenter during one annual revolution of each planet (Fig. 2).

The ratio of the amplitudes of the orbit-spin coupling torques on the two bodies is somewhat larger than the factor of 3 or so indicated in the forcing function plots of Fig. A1. If we wish to compare accelerations at the surfaces of these planets, due to the factor *r* in equation (3), we must additionally account for the differences in the planetary dimensions. The ratio of the planetary radii is ~1.88 (6.378 km / 3.396 km).

Peak values of the orbit-spin coupling accelerations at the surface of Mars for the period 1920-2030 were calculated in Mischna & Shirley 2017). An acceleration of $2.2 \times 10^{-4}$ m s$^{-2}$ was obtained when employing a value for the coefficient *c* of $5 \times 10^{-13}$. A similar calculation, made for the time of the largest positive peak of the terrestrial *dL/dt* waveform in late 2020, is provided above in Section 2.3. That calculation returns a value of $1.2 \times 10^{-3}$ m s$^{-2}$, when likewise employing a *c* value of $5 \times 10^{-13}$. Thus our current best estimate is that the *cta* accelerations and the torque on the Earth are likely to be larger than the corresponding Martian values by a factor of ~5.

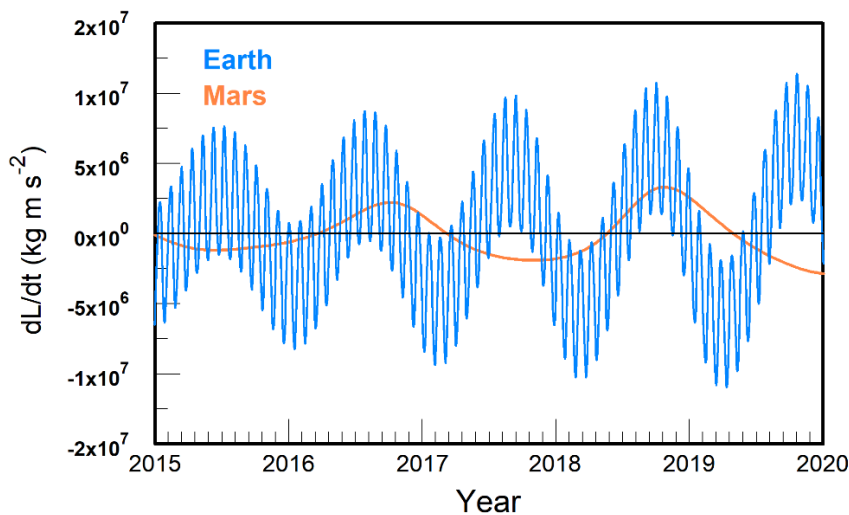

**FIG. A1**. Comparison of rates of change of orbital angular momenta (*dL/dt*) for Earth and Mars (with respect to the solar system barycenter) for the years 2015-2020. The figure plots values of the *z* component of *dL/dt* with respect to ecliptic coordinates. Units are as in Fig. 3.

**Appendix 2: Considerations impacting the selection of time intervals for investigation**

While some of the drought atlases now available (Baek et al., 2015) extend back more than two millennia, we chose to restrict the time interval for this investigation to the period of available instrumental observations (Table 2). The reasons for this choice include the following. First, we recognize the increasing difficulty of defining accurate start and ending dates for multiyear drought episodes occurring in earlier centuries. Data quality for earlier times may be inferior to that for the instrumental period, necessarily introducing additional uncertainty to the analysis. Secondly, given the preliminary nature of the present investigation, we considered that the selected 180-yr time span would be sufficient to ensure an adequate test of the underlying hypothesis. Third, we further recognize that the analysis time requirement may be expected to increase roughly in proportion to the volume of Earth system torque data employed.

Possible heterogeneities of Earth system response over longer time periods comprise one further reason to limit the analysis to the most recent period. The climates of the Little Ice Age (from ~1550-1830 CE) (Lamb, 1977) were significantly different from those of the modern period, as were those of Medieval times (~800-1300 CE). While the physics of orbit-spin coupling would be unchanged, the Earth system response during these intervals might differ significantly from that of recent times, due for instance to likely differences in SSTs.

That said, expanding the analysis to earlier times is a worthwhile objective for later investigations. The Medieval period is of particular interest, due to the occurrence of several megadroughts of significantly longer duration than those listed here in Table 2 (Cook et al., 2004, 2010; Woodhouse et al., 2010). The impacts on societies and the environment associated with droughts lasting decades would be more devastating than those of recent times.

From a solar system dynamics perspective: We note in passing that the 5-body system of the Sun and the giant planets exhibits millennial time scale variability. The dynamics of this 5-body system control the time variability of the EMB torque component described here in § 4.1. Significantly reduced time variability of a dynamical parameter characterizing the Sun's motion during the Medieval period has been recognized (cf. Fig. 5 of Fairbridge & Shirley (1987). Further work is needed to better understand the relationships between solar system dynamical forcing and the Earth system response in Medieval times.

Loder, J. W., and Garrett, C., 1978. The 18.6-year cycle of sea surface temperature in shallow seas due to variations in tidal mixing, Journal of Geophysical Research 83, 1967-1970.

Lotsch, A., Friel, M. A., Anderson, B. T., and Tucker, C. J., 2005. Response of terrestrial ecosystems to recent Northern Hemispheric drought, Geophysical Research Letters 32, L06705, doi:10.1029/2004GL022043

Lorentz, E. N., 1963. Deterministic Non-periodical Flow. *Journal of the Atmospheric Sciences* 20, 130-141.

Lorentz, E. N., 1969. Atmospheric Predictability as Revealed by Naturally Occurring Analogs, *Journal of the Atmospheric Sciences* 26, 636-646.

Lorentz, E. N., 1984. Irregularity: A Fundamental Property of the Atmosphere, *Tellus* 36A, 98-110.

MacKinnon, J. A., Zhao, Z., Whalen, C. B., Waterhouse, A. F., Trossman, D. S., et al., 2017. Climate process team on internal wave-driven ocean mixing, Bulletin of the American Meteorological Society (November 2017), 2429-2454, doi:10.1175/BAMS-D-16-0030.1

Mantua, N. J., Hare, S. R., Zhang, Y., Wallace, J. M., and Francis, R. C., 1997. A Pacific interdecadal climate oscillation with impacts on salmon production, Bulletin of the American Meteorological Society 78, 1069-1079.

Mariotti, A., Ruti, P. M., and Rixen, M., 2018. Progress in subseasonal to seasonal prediction through a joint weather and climate community effort, npj Climate & Atmospheric Science 1, 4, https://doi.org/10.1038/s41612-018-0014-z

Martin, J. T., and 18 others, 2020. Increased drought severity tracks warming in the United States' largest river basin, Proceedings of the National Academy of Sciences 117, 11328-11336, 10.1073/pnas.1916208117

Mazzarella, A., and Palumbo, A., 1994. The lunar nodal induced signal in climatic and oceanic data over the Western Mediterranean Area and on its bistable phasing, Theoretical and Applied Climatology 50, 93-102.

McKinnell, S. M., and Crawford, W. R., 2007. The 18.6-year lunar nodal cycle and surface temperature variability in the northeast Pacific, Journal of Geophysical Research 112, C02002, doi:10.1029/2006JC003671

McPhaden, M. J., Zebiak, S. E., and Glantz, M. H., 2006. ENSO as an integrating concept in Earth science, Science 314, 1740-1745.

Meehl, G. A., Goddard, L., Boer, G., Burgman, R., Branstator, G., et al. 2014. Decadal climate prediction, Bulletin of the American Meteorological Society (February 2014), 243-267, doi:10.1175/BAMS-D-12-00241.1

Teng, H. and Branstator, G., 2017. Causes of extreme ridges that induce California droughts, Journal of Climate 30, 1477-1492, doi:10.1175/JCLI-D-16-0524.1

Toigo, A.D., Lee, C., Newman, C.E., Richardson, M.I., 2012. The impact of resolution on the dynamics of the martian global atmosphere: varying resolution studies with the MarsWRF GCM. Icarus 221, 276–288.

Trenberth, K. E., Branstator, G. W., and Arkin, P. A., 1988. Origins of the 1988 North American drought, Science 242, 1640-1645.

Trenberth, K. E., Dai, A., van der Schrier, G., Jones, P. D., Barichivich, J., Briffa, K., and Sheffield, J., 2013. Global warming and changes in drought, Nature Climate Change 4, 17-22, doi:10.1038/nclimate2067

Tyson, P. D., Dyer, T. G. J., and Mametse, M. N., 1975. Secular changes in South African rainfall: 1880 to 1972, Quarterly Journal of the Royal Meteorological Society 101, 817-833.

Vines, R. G., 1982. Rainfall Patterns in the Western United States. *J. Geophys. Res*. 87, 7303-7311.

Vines, R. G., 1985. Rainfall patterns in Europe, Journal of Climatology 5, 607-616.

Vines, R. G., 1986. Rainfall patterns in India, Journal of Climatology 6, 135-148.

Visage, P. J., 1966. Preciptation in South Africa and lunar phase, Journal of Geophysical Research 71, 3345-3350.

Visvanathan, T. R., 1966. Formation of depressions in the Indian seas and lunar phase, Nature 210, 406-407.

Wang, S.-Y., Hipps, L., Gillies, R. R., and Yoon, J.-H., 2014. Probable causes of the abnormal ridge accompanying the 2013-2014 California drought: ENSO precursor and anthropogenic warming footprint, Geophysical Research Letters 41, 3220-3226, doi:10.1002/2014GL059748

Wells, N.C., 2007. The Atmosphere and Ocean: A Physical Introduction. (§1.3-1.4), (448 pps), Wiley.

Willhite, D. A., 2000. Droughts as a natural hazard: Concepts and definitions, in Drought: A Global Assessment, Ed. Willhite, D. A., Hazards and Disaster Series (Volume 2), pp. 3-18, Routledge.

Williams, A. P., Seager, R., Abatzoglou, J. T., Cook, B. I., Smerdon, J. E., and Cook, E. R., 2015. Contribution of anthropogenic warming to California drought during 2012-2014, Geophysical Research Letters 42, 6819-6828, doi:10.1002/2015GL064924

Wills, R. C. J., Battisti, D. S., Proistosescu, C., Thompson, L., Hartmann, D. L., and Armour, K. C., 2019. Ocean circulation signatures of North Pacific decadal variability, Geophysical Research Letters 46, 1690-1701, doi:10.1029/2018GL080716
65

**Acknowledgements and Data**